\documentclass[a4paper,11pt]{article}
\pdfoutput=1 % if your are submitting a pdflatex (i.e. if you have
\date{}
\usepackage{jheppub}
\usepackage{amsmath,amssymb,amsfonts,amsxtra, mathrsfs,graphics,graphicx,amsthm,epsfig, youngtab,bm,longtable,float,tikz,empheq}
\usepackage[margin=0.5cm]{caption}

\newcommand{\CN}{\mathcal{N}}

\newcommand {\apgt} {\ {\raise-.5ex\hbox{$\buildrel>\over\sim$}}\ }
\newcommand {\aplt} {\ {\raise-.5ex\hbox{$\buildrel<\over\sim$}}\ }

\makeatletter\@addtoreset{equation}{section}\makeatother

\def\Z{\relax\ifmmode\mathchoice
{\hbox{\cmss Z\kern-.4em Z}}{\hbox{\cmss Z\kern-.4em Z}} {\lower.9pt\hbox{\cmsss Z\kern-.4em Z}}
{\lower1.2pt\hbox{\cmsss Z\kern-.4em Z}}\else{\cmss Z\kern-.4em Z}\fi}

\def\e{{\epsilon}}

\def\btimes{~{{{\lower1pt\hbox{$\square$}} \kern-7.6pt \times}}~}

\def\Z{{\Bbb{Z}}}

\def\be{\begin{equation}}
\def\ee{\end{equation}}
\newcommand{\bea}{\begin{eqnarray}}
\newcommand{\eea}{\end{eqnarray}}

\renewcommand{\bar}{\overline}
\renewcommand{\hat}{\widehat}
\renewcommand{\tilde}{\widetilde}

\newcommand{\Reals}{\mathbb{R}}
\newcommand{\Complex}{\mathbb{C}}

\makeatletter
\newcommand{\ellSN}{\mathop{\operator@font sn}\nolimits}
\newcommand{\ellCN}{\mathop{\operator@font cn}\nolimits}
\newcommand{\ellDN}{\mathop{\operator@font dn}\nolimits}
\newcommand{\ellAM}{\mathop{\operator@font am}\nolimits}
\newcommand{\ellK}{\mathop{\smash{\operator@font K}\vphantom{a}}\nolimits}
\newcommand{\ellE}{\mathop{\smash{\operator@font E}\vphantom{a}}\nolimits}
\makeatother

\ifx\genfrac\sdflkaj

\else

\fi

\newcommand{\beq}{\begin{equation}}
\newcommand{\eeq}{\end{equation}}

\makeatletter
\def\mr@ignsp#1 {\ifx\:#1\@empty\else #1\expandafter\mr@ignsp\fi}%
\newcommand{\multiref}[1]{\begingroup%\let\protect\string%
\xdef\mr@no@sparg{\expandafter\mr@ignsp#1 \: }%
\def\mr@comma{}%
\@for\mr@refs:=\mr@no@sparg\do{\mr@comma\def\mr@comma{,}\ref{\mr@refs}}%
\endgroup}
\makeatother

\newcommand{\hypref}[2]{\ifx\href\asklfhas #2\else\href{#1}{#2}\fi}
\newcommand{\Secref}[1]{Section~\multiref{#1}}
\newcommand{\secref}[1]{Sec.~\multiref{#1}}

\renewcommand{\eqref}[1]{(\multiref{#1})}

\def\[{\begin{equation}}
\def\]{\end{equation}}
\def\<{\begin{eqnarray}}
\def\>{\end{eqnarray}}

\ifx\href\asklfhas\newcommand{\href}[2]{#2}\fi

\title{On Elliptic Algebras and Large-$n$ Supersymmetric Gauge Theories}

\author[a]{Peter Koroteev,}
\author[b]{Antonio Sciarappa}

\affiliation[a]{Perimeter Institute for Theoretical Physics \\ 31 Caroline Street North, Waterloo, ON N2L2Y5, Canada}
\affiliation[a]{Department of Mathematics, University of California \\ One Shields Avenue, Davis, CA 95616, USA}
\affiliation[b]{School of Physics, Korea Institute for Advanced Study, \\Seoul 130-722, Korea}

\emailAdd{pkoroteev@perimeterinstitute.ca}
\emailAdd{asciara@kias.re.kr}

\abstract{In this note we further develop the duality between supersymmetric gauge theories in various dimensions and elliptic integrable systems such as Ruijsenaars-Schneider model and periodic intermediate long wave hydrodynamics. These models arise in instanton counting problems and are described by certain elliptic algebras. We discuss the correspondence between the two types of models by employing the large-$n$ limit of the dual gauge theory. In particular we provide non-Abelian generalization of our previous result on the intermediate long wave model.}

\preprint{KIAS-P16010}

\begin{document}

\maketitle

\section{Introduction} \label{Sec:Intro}
In physics literature there is a significant amount of interest towards gauge theories with large number of colors (large-$n$\footnote{We deliberately denote number of colors with lower-case $n$ in order to be consistent with some literature on integrable systems} gauge theories). One of the most commonly utilized benefits of the large-$n$ physics is the simplification of perturbative expansions, where a multitude of Feynman diagrams is $1/n$ suppressed \cite{Hooft:1973jz}. However, the non-perturbative phenomena of large-$n$ gauge theories are discussed less frequently. Indeed, in a generic setup they are shadowed by the perturbative contributions which scale with $n$. Nevertheless there are some notable exceptions when non-perturbative objects, such as instantons, do play an important r\^ole in large-$n$ physics. Our current work will elaborate on one of these possibilities.
One typically expects to construct an effective description of a gauge theory in the large-$n$ limit which often involves different degrees of freedom than those of the original theory. The effective theory usually appears to be more tractable and sometimes exactly soluble. Expectedly, when the gauge theory is supersymmetric, there are more grounds for deeper understanding of the effective theory by studying its BPS sector and the large-$n$ limit of its protected observables.

In this paper we shall investigate an effective large-$n$ description of a certain $U(n)$ gauge theory of $\hat{A}_0$-type with eight supercharges in five dimensions\footnote{The theory can thought of as a reduction of the 6d $(2,0)$ theory on a circle of a radius which dials the five-dimensional gauge coupling. It is believed that the theory is UV complete if one includes instantons and tensionless monopole strings (see e.g. \cite{Kim:2011aa} and references therin). The BPS observables which will be computed in this work are insensitive to the details of the UV completion.} on $\Reals^4\times S^1$ thereby extending our previous work \cite{Koroteev:2015dja}. In \textit{loc cit} we have shown that the vacuum expectation value of the fundamental Wilson loop wrapped around $S^1$ has a well-defined infinite-$n$ limit where it reproduces a different observable of three-dimensional quiver gauge theory with four supercharges. The latter quiver serves in the ADHM construction \cite{Atiyah:1978ri} of the moduli space of (non-commutative) $U(1)$ instantons \cite{Douglas:xy}. Therefore we have constructed a correspondence between the 5d instanton counting and instantons of a completely different theory. In other words, the non-perturbative effects of the original gauge theory survived the large-$n$ limit and provided us with a novel effective description. As explained in \cite{Koroteev:2015dja} there is a nontrivial matching of the parameters of both theories which has certain physical and mathematical implications. \Secref{Sec:sec2} of the present paper is entirely devoted to brief review of \cite{Koroteev:2015dja} such that it would prepare the reader for the new results which we have derived by studying a \textit{different} limit of the 5d $\CN=1^*$ theory\footnote{All our results apply to four dimensional theories with eight supercharges as well; the reason for us to stay in five dimensions is rather technical. There are certain advantages from both physics and the representation theory point of view (see \cite{Koroteev:2015dja} for details).}.

Remarkably the above paragraph can be reinterpreted using the language of integrable systems. It has been known that infrared physics of gauge theories with eight supercharges, which is elegantly described by Seiberg-Witten solution \cite{Seiberg:1994aj,Seiberg:1994rs}, has an equivalent presentation in terms of classical exactly soluble many-body systems \cite{Gorsky:1995zq, Donagi:1995cf,Martinec:1995by}. In the past several years there has been a significant progress in understanding quantization of these models using gauge theories in Omega background \cite{Nekrasov:2009rc} and, more recently, by studying moduli spaces of instantons with ramification \cite{Alday:2010vg,Bullimore:2015fr,Nekrasov:2015wsu}. The latter approach can be realized by adding codimension-two (or monodromy-type \cite{Gukov:2006jk}) defects on the worldvolume of the gauge theory\footnote{Also codimension-two defect on M5 branes which engineer the gauge theory if it is of class $\mathcal{S}$.} which supplement the gauge bundle by adding to it several nontrivial first Chern classes -- one for each monodromy parameter. Thus for each gauge theory one can assign a quantum Hamiltonian which acts on the space of the monodromy parameters of the defect. We can then ask what are the eigenvalues and the eigenstates of this Hamiltonian. According to the gauge/inegrability dictionary if we start with the 5d $\CN=1^*$ theory on $\Reals^4\times S^1$ with gauge group $U(n)$, then the corresponding integrable system is the $n$-body elliptic Ruijsenaars-Schneider model \cite{MR1329481,MR1322943,MR929148,MR887995,MR851627}. As it was shown in 
\cite{Bullimore:2015fr} the quantum Ruijsenaars-Schneider operator has the following formal solution -- its eigenfunctions are supersymmetric partition functions of the 5d theory in the presence of the monodromy defect of a maximal Levi type, whereas its eigenvalues are vacuum expectations values of the Wilson loop (in different skew-symmetric powers of the fundamental representation) wrapping the compact circle (cf. previous paragraph). Therefore we can reformulate the stable limit of the instanton configurations which we have discussed earlier in terms of the spectrum of the elliptic Ruijsenaars-Schneider (eRS) model with infinite number of particles. 

As we explained in \cite{Koroteev:2015dja} the effective large-$n$ description of the elliptic Ruijsenaars-Schneider model is the finite difference intermediate long wave hydrodynamical system ($\Delta$ILW) \cite{2009JPhA...42N4018S,2009arXiv0911.5005T,2011ntqi.conf..357S}. This is a hydrodynamical system which is described by a certain difference-integral equation for a velocity field of a fluid in one-dimensional periodic channel. It is known to be integrable and its spectrum can be mapped onto the twisted chiral ring of the ADHM quiver which we mentioned above. In particular, the generating function of the $\Delta$ILW spectrum coincides with one of the Casimirs of the vectormultiplet scalar of the ADHM theory. Mathematically $\Delta$ILW Hamiltonians enter the Fock space representation of elliptic Ding-Iohara algebra \cite{Ding:1996mq} which are deeply connected with the elliptic Ruijsenaars-Schneider model. Plethora of limiting cases from both eRS and $\Delta$ILW models, which describe certain physical regimes including Calogero-Moser/ILW, Benjamin-Ono, etc. was earlier studied in the literature (see \cite{Koroteev:2015dja} for details).

The Hamiltonians of the elliptic Ruijsenaars-Schneider model model can be thought of as certain elliptic generalizations of Macdonald operators \cite{MR553598}, and its eigenfunctions serve as series generalizations of Macdonald polynomials. Macdonald operators appear in representations of double affine Hecke algebras \cite{9780511546501} which in turn describe Hamiltonians of the trigonometric Ruijsenaars-Schneider model \cite{Bullimore:2015fr}. In \cite{Koroteev:2015dja} we discussed free field realization of Macdonald polynomials which can be realized via Ding-Iohara algebra \cite{Ding:1996mq, 1995q.alg.....7034S}. Therefore free boson presentation of the eRS model calls for an elliptic generalization of the algebra. Presently in the literature there are two distinct generalizations. First generalization presented by Feigin et al \cite{Feigin:2009ab}, which was used in \cite{Koroteev:2015dja} and will be employed in the current work, and, second, the one by Saito \cite{2013arXiv1301.4912S} which leads to the so-called elliptic Virasoro algebra recently studied in \cite{Nieri:2015dts,Iqbal:2015fvd}. At the moment it is not known how to relate the two approaches, however, we believe that they should be in some sense identical.

There is a peculiar non-Abelian generalization of the ILW system, which is referred to as ILW$_N$. Morally speaking it represents a fluid with non-Abelian velocity fields $u^{a}(t,x),\, a=1,\dots, N$ which interact with each other in a way that respects the $U(N)$ invariance. Presently in the literature not much is known about the difference version of the intermediate long wave system, which we call $\Delta$ILW$_N$, however there are some results for ILW$_N$ \cite{lebedev1983,ref1,2013JHEP...11..155L,2015JHEP...02..150A} and its Benjamin-Ono limit \cite{2005PhRvL..95g6402A,2009JPhA...42m5201A}. Nevertheless, using the relationship with supersymmetric gauge theories, we will be able to predict its spectrum, in particular we shall use its relation to the moduli space of $U(N)$ instantons (this will be done in \Secref{Sec:sec4}). Finally we will be able to demonstrate that the $\Delta$ILW$_N$ arises as a certain $n\to\infty$ limit of the 5d $U(Nn)$ $\CN=1^*$ gauge theory and provide a direct mapping between the parameters of both systems in \Secref{Sec:sec5}. Later in \Secref{quantumcohomology} we discuss the relationship of $\Delta$ILW$_N$ Hamiltonians and quantum multiplication in quantum cohomology ring of $\mathcal{M}_{k,N}$.

\section{Review of the eRS/$\Delta$ILW Correspondence} \label{Sec:sec2}
In this section we shall review the derivation found in \cite{Koroteev:2015dja} of the duality between the elliptic Ruijsenaars-Schneider system and the difference ILW model. We first discuss the trigonometric limit, or the tRS/$\Delta$BO duality, then we shall address free field representation of the Ruijsenaars-Schneider systems followed by the large-$n$ limit and connections to $\Delta$ILW$_1$.

\subsection{Trigonometric and Elliptic Ruijsenaars Systems} \label{Sec:sec2.1}
The $n$-particle trigonometric Ruijsenaars-Schneider model\footnote{In this paper we are considering the complexified system, which means that all coordinates and momenta are considered to be complex. By imposing appropriate reality conditions on the parameters, we can reproduce the real trigonometric system or the real hyperbolic one.} (tRS) is a complex quantum integrable system of $n$ interacting particles living on a cylinder. The dynamics is determined by the Hamiltonian 
\begin{equation}
D^{(1)}_{n,\vec{\tau}}(q,t) = \sum_{i=1}^n \prod_{j \neq i}^n \dfrac{t \tau_i - \tau_j}{\tau_i - \tau_j} T_{q,i} \,, \label{tRSh1}
\end{equation}
which is also the first conserved quantity of the integrable system; the whole set of $n$ conserved quantities is given by
\begin{equation}
D_{n,\vec{\tau}}^{(r)}(q,t) = t^{r(r-1)/2} \sum_{\substack{I \subset \{1,2,\ldots, n\} \\ \# I = r}} \prod_{\substack{ i \in I \\ j \notin I}} \dfrac{t \tau_i - \tau_j}{\tau_i - \tau_j} \prod_{i \in I} T_{q,i} \;\;\; \text{for} \;\;\; r = 1, \ldots, n \,. \label{tRShn}
\end{equation}
The meaning of the parameters is the following: $\tau_i$ are positions of the particles, $t$ is interaction coupling, and $T_{q,i}$ is shift operator acting as
\begin{equation}
T_{q,i} f(\tau_1, \ldots, \tau_i, \ldots, \tau_n) = f(\tau_1, \ldots, q \tau_i, \ldots, \tau_n)
\end{equation}
on functions of the $\tau_i$ variables. This operator can be represented as $T_{q,i} = e^{i \gamma \epsilon_1 \tau_i \partial_{\tau_i}} = q^{\tau_i \partial_{\tau_i}}$, where $\gamma$ is related to the radius of the circle of the cylinder and $\epsilon_1$ plays the role of the Planck constant $\hbar$; in fact, this is merely a trigonometric version of the usual quantum momentum operator. 

Macdonald polynomials $P_{\lambda}(\vec{\tau};q,t)$ are eigenfunctions of the tRS system. These are symmetric polynomials of degree $k$ in the $n$ variables $\tau_i$, and are in one-to-one correspondence with partitions $\lambda = (\lambda_1, \ldots, \lambda_n)$, $\lambda_1 \geqslant \ldots \geqslant \lambda_n \geqslant 0$ of $k$ of length $n$. Being symmetric, they can be written as linear combinations of the power-sum polynomials 
\begin{equation}
p_m = \sum_{i=1}^n \tau_i^m\,.
\end{equation}
An eigenfunctions $P_{\lambda}(\vec{\tau};q,t)$ satisfy\footnote{As remarked for example in \cite{Zenkevich:2014lca}, for generic $q,t$ the spectrum of $D_{n,\vec{\tau}}^{(1)}(q,t)$ is non-degenerate, so there is no need for considering the higher Hamiltonians $D_{n,\vec{\tau}}^{(r)}(q,t)$ to completely classify the eigenfunctions. This is one of the reasons why we will often not consider the whole set of Hamiltonians in the following.}
\begin{equation}
D_{n,\vec{\tau}}^{(1)}(q,t) P_{\lambda}(\vec{\tau}; q, t) = E_{tRS}^{(\lambda; n)} P_{\lambda}(\vec{\tau}; q, t) 
\end{equation}
with eigenvalue given by
\begin{equation}
E_{tRS}^{(\lambda; n)} = \sum_{j=1}^n q^{\lambda_j} t^{n-j} \,.
\end{equation}
As it is clear from this expression, the partition $\lambda$ completely determines the eigenvalue. \\

\noindent The tRS system admits an elliptic generalization, known as the elliptic Ruijsenaars-Schneider (eRS) system, which consists of $n$ particles on a torus, whose dynamics is determined by Hamiltonian
\begin{equation}
D_{n,\vec{\tau}}^{(1)}(q,t;p) = \sum_{i=1}^n \prod_{j \neq i}^n \dfrac{\Theta_p(t \tau_i/\tau_j)}{\Theta_p(\tau_i/\tau_j)} T_{q,i} \,. \label{eRSh1}
\end{equation}
Here $\Theta_p(x)$ is defined as
\begin{equation}
\Theta_p(x) = (p;p)_{\infty} (x;p)_{\infty} (p/x;p)_{\infty}\,, \quad (x;p)_{\infty} = \prod_{s=0}^{\infty} (1-x p^s)\,.
\end{equation}
In the limit $p \rightarrow 0$ the elliptic curve where the above theta functions are defined degenerates into a cylinder and \eqref{eRSh1} reduces to \eqref{tRSh1}. Eigenfunctions and eigenvalues of this model are not known in closed form; it is nevertheless possible to obtain them as a series expansion in $p$ around the known tRS solutions\footnote{See \cite{2004math.ph...7050L} for an analogous treatment of solutions of the elliptic Calogero model, also known as the non-relativistic limit of the eRS system}. This approach for determining the solution of the eRS system is the one followed in \cite{Bullimore:2015fr}, although the techniques used there come from 5d supersymmetric gauge theories. Nekrasov recently proved using a different method \cite{Nekrasov:2015wsu} that the proposal of \cite{Bullimore:2015fr} provides a solution for the elliptic Calogero model.

According to the correspondence between integrable many-body systems and supersymmetric gauge theories the $n$-particle eRS system can be realized in terms of a 5d $\mathcal{N}=1^*$ $U(n)$ theory in the Omega background $\mathbb{C}^2_{\epsilon_1, \epsilon_2} \times S^1_{\gamma}$ in the presence of codimension-two defects. Codimension-two defects correspond to a 3d $T[U(n)]$ theory living on $\mathbb{C}_{\epsilon_1} \times S^1_{\gamma}$; the coupled 5d/3d instanton partition function $Z^{\text{inst}}_{5d/3d}$ in the Nekrasov-Shatashvili limit $\epsilon_2 \rightarrow 0$ \cite{Nekrasov:2009rc} is a formal eigenfunction of the eRS system. On the other hand, the vacuum expectation values of the Wilson loop $\left\langle W_{\square}^{U(n)} \right\rangle$ in the fundamental representation of $U(n)$ gives the eigenvalues of the eRS system, again in the Nekrasov-Shatashvili limit. The 5d instanton parameter $Q = e^{-8\pi^2 \gamma/g_{YM}^2}$ is identified with the eRS elliptic deformation parameter $p$. When the 5d gauge coupling is turned off the 5d theory decouples, leaving us with purely three-dimensinoal theory, which in turn is dual to the trigonometric Ruijsenaars-Schneider system. We refer to \cite{Bullimore:2015fr} (see also \cite{Koroteev:2015dja}) for further details and the precise dictionary between gauge theory and eRS parameters. 

Let us stress that the solution provided by gauge theory computations is a \textit{formal} solution, viz. its eigenfunctions might not be normalizable. Moreover, already at the trigonometric level it looks quite different from the tRS solution discussed above, since both eigenfunctions and eigenvalues depend on the 5d Coulomb branch parameters $\mu_a$, $a = 1, \ldots, n$. In fact, as is noted in \cite{Koroteev:2015dja}, both problems can be cured by noticing that if we set
\begin{equation}
\mu_a = q^{\lambda_a}t^{n-a} \;\;\;,\;\;\; a = 1, \ldots, n \label{locus}
\end{equation}
for a given partition $\lambda$ of an integer $k$, than the formal eigenfunctions reduce to Macdonald polynomials associated to the corresponding partition at the trigonometric level, while they become symmetric polynomials in the ratios $\frac{\tau_i}{\tau_j}$ of coordinates when the elliptic parameter is turned on. In this way the eigenfunctions become normalizable with the standard Macdonald measure, and we recover the usual tRS solution in the trigonometric limit. Equation \eqref{locus} specifies the very special locus in the Coulomb branch of our 5d theory in which a Higgs branch opens up `Higgs branch root', and vortex strings may emerge \cite{Chen:2012we}. 

Taking \eqref{locus} into account, we can now make a proper use of the gauge theory computations relative to the eRS system. In particular, in the following we will focus on the eigenvalue $E_{eRS}^{(\lambda;n)}(p)$ of the first eRS Hamiltonian \eqref{eRSh1} relative to an eigenfunction labelled by a partition $\lambda$, which according to gauge theory is given by \cite{Bullimore:2015fr,Koroteev:2015dja}
\begin{equation}
E_{eRS}^{(\lambda;n)}(p) \;=\; \left\langle W_{\square}^{SU(n)} \right\rangle\Big\vert_{\lambda} \;=\; \left\langle W_{\square}^{U(n)} \right\rangle \Big/ \left\langle W_{\square}^{U(1)} \right\rangle\Big\vert_{\lambda} \,,\label{WSU}
\end{equation}
where
\begin{equation}
\left\langle W_{\square}^{U(n)} \right\rangle = \sum_{a=1}^n \mu_a - Q \dfrac{(q-t)(1-t)}{q t^n} \sum_{a=1}^n \mu_a
\prod_{\substack{b=1 \\ b \neq a}}^n \dfrac{(\mu_a - t \mu_b)(t \mu_a - q \mu_b)}{(\mu_a - \mu_b)(\mu_a - q \mu_b)} + o(Q^2)\,,
\end{equation}
\begin{equation}
\left\langle W_{\square}^{U(1)} \right\rangle = \dfrac{(Qt^{-1};Q)_{\infty}(Qtq^{-1};Q)_{\infty}}{(Q;Q)_{\infty}(Qq^{-1};Q)_{\infty}}\,.
\end{equation}
Formula \eqref{WSU} will play an important r\^ole in the following discussion as well as it did in \cite{Koroteev:2015dja}: there it was used to show that, in the limit of large $n$, the eRS model can be described in terms of a quantum hydrodynamic system known as finite-difference Intermediate Long Wave system ($\Delta$ILW), or finite-difference Benjamin-Ono ($\Delta$BO) in the trigonometric case. This correspondence to hydrodynamic models is easy to understand at the classical level -- when a system consists of an infinite number of particles it is impossible to follow the dynamics of every single particle, and a better description of the system can be provided by considering it as a fluid, i.e. by studying the particles' collective motion. This idea can be translated at the quantum language -- one now needs to expand the fluid velocity functions in Fourier modes and then quantize these modes according to the canonical quantization procedure. This is equivalent to consider our original eRS system in its (bosonic) \textit{free field} (or \textit{collective field}) \textit{representation} \cite{Jevicki:1979mb,Minahan:1994hn,1995PhLB..352..111I} (see \cite{1995PhLB..347...49A,2000math.ph...7036L,2004CMaPh.247..321L} for the collective field description of trigonometric and elliptic Calogero-Sutherland systems). Further details on this approach for the case at hand can be found in \cite{Koroteev:2015dja}; in the next subsection we will merely collect some basic facts which will be relevant for the upcoming discussion.

\subsection{Free Field Realization of Ruijsenaars-Schneider Systems} \label{Sec:sec2.2}
The free field realization of tRS and eRS models has been discussed in great detail in \cite{Feigin:2009ab} (see also \cite{2013arXiv1301.4912S,2014SIGMA..10..021S,2013arXiv1309.7094S} for a different realization). We start by considering the $(q,t)$-deformed Heisenberg algebra $\mathcal{H}(q,t)$, generated by the $a_m$, $m \in \mathbb{Z}$ modes following the commutation relation
\begin{equation}
[a_m,a_n] = m \dfrac{1-q^{\vert m \vert}}{1-t^{\vert m \vert}} \delta_{m+n,0}\,. 
\end{equation}
In order to reproduce the action of the first trigonometric Ruijsenaars-Schneider Hamiltonian \eqref{tRSh1} in terms of Heisenberg modes $a_m$ we introduce vertex operators
\begin{equation}
\begin{split}
\eta(z) & \,=\, \text{exp} \left( \sum_{n>0} \dfrac{1-t^{-n}}{n} a_{-n}z^n \right) \text{exp} \left( - \sum_{n>0} \dfrac{1-t^n}{n} a_n z^{-n} \right) \\
& \,=\, :\text{exp}\left( - \sum_{n \neq 0} \dfrac{1-t^n}{n} a_n z^{-n} \right): \,=\, \sum_{n \in \mathbb{Z}} \eta_{n}z^{-n} \label{etat}
\end{split}
\end{equation}
and
\begin{equation}
\phi(z) = \text{exp} \left( \sum_{n>0} \dfrac{1-t^n}{1-q^n}a_{-n}\dfrac{z^n}{n} \right) \,;
\end{equation}
now, after defining $\phi_n(\tau) = \prod_{i=1}^n \phi(\tau_i)$, one can show that \cite{Feigin:2009ab}
\begin{equation}
\mathcal{O}_1(q,t) \phi_n(\tau) \vert 0 \rangle \equiv [\eta(z)]_1 \phi_n(\tau) \vert 0 \rangle = \left[ t^{-n} + t^{-n+1}(1-t^{-1})D^{(1)}_{n,\vec{\tau}}(q,t) \right] \phi_n(\tau) \vert 0 \rangle \label{keytr}
\end{equation}
where $[\;\;]_1$ means the constant term in $z$ (i.e. $[\eta(z)]_1 = \eta_0$). At the level of eigenvalues this means that for a fixed eigenstate labelled by a partition $\lambda$ we have
\begin{equation}
\mathcal{E}_{1;(\lambda)} = t^{-n} + t^{-n+1}(1-t^{-1})E^{(\lambda; n)}_{tRS}\,, \label{trigeig}
\end{equation} 
where $\mathcal{E}_{1}^{(\lambda)}$ is the eigenvalue of the $[\eta(z)]_1$ operator. This implies for $\vert t \vert < 1$
\begin{equation}
\mathcal{E}_{1}^{(\lambda)} = \lim_{n \rightarrow \infty} \left[ t^{-n+1}(1-t^{-1}) E_{tRS}^{(\lambda; n)} \right] 
\end{equation}
as can be easily verified \cite{Koroteev:2015dja}. Let us mention here that the eigenfunctions of $[\eta(z)]_1$ can be easily obtained from the Macdonald polynomials written in terms of the power sum polynomials $p_m$ thanks to the isomorphism between the space of symmetric polynomials and the Fock space vectors of $\mathcal{H}(q,t)$ given by\footnote{In the elliptic case this isomorphism will no longer be of help, since we need to consider symmetric polynomials in the ratios $\frac{\tau_i}{\tau_j}$ which cannot be written as linear combinations of $p_m$.}
\begin{equation}
a_{-m} \vert 0 \rangle \;\; \longleftrightarrow \;\; p_m \,. \label{isomorphism}
\end{equation}
In a similar way, the action of the higher order Hamiltonians \eqref{tRShn} can be expressed in terms of bosonic oscillators through commuting operators $\mathcal{O}_r(q,t)$ ($r = 1, \ldots, n$) which are constructed out of the normal ordered product of $r$ vertex operators $\eta(z_i)$, $i = 1, \ldots, r$. When we consider the tRS system in the limit $n \rightarrow \infty$ we therefore obtain an infinite set of commuting quantum operators $\mathcal{O}_r(q,t)$, $r = 1, \ldots, \infty$.  In \cite{Koroteev:2015dja} these have been proposed to be the Hamiltonians defining the quantum $\Delta$BO hydrodynamic system, based on the analysis of the classical $\Delta$BO system of \cite{2009arXiv0911.5005T,2011ntqi.conf..357S}. 

The same procedure can be adopted for the eRS model. We simply need to replace \eqref{etat} by
\begin{equation}
\eta(z;pq^{-1}t) = \text{exp} \left( \sum_{n>0} \dfrac{1-t^{-n}}{n} \dfrac{1-(pq^{-1}t)^n}{1-p^n} a_{-n}z^n \right) \text{exp} \left( -\sum_{n>0} \dfrac{1-t^n}{n} a_n z^{-n} \right)\,, \label{etap}
\end{equation}
with parameter of elliptic deformation $p$. Equation \eqref{keytr} gets modified into
\begin{equation}
\begin{split}
& \mathcal{O}_1(q,t;p) \phi_n(\tau;p) \equiv \left[ \eta(z;pq^{-1}t) \right]_1 \phi_n(\tau;p) \vert 0 \rangle = \\
& \phi_n(\tau;p) \left[ t^{-n} \prod_{i=1}^n \dfrac{\Theta_p(qt^{-1}z/\tau_i)}{\Theta_p(qz/\tau_i)} \dfrac{\Theta_p(tz/\tau_i)}{\Theta_p(z/\tau_i)} \eta(z;pq^{-1}t) \right]_1 \vert 0 \rangle \\
& + t^{-n+1}(1-t^{-1})\dfrac{(pt^{-1};p)_{\infty}(ptq^{-1};p)_{\infty}}{(p;p)_{\infty}(pq^{-1};p)_{\infty}} D^{(1)}_{n,\vec{\tau}}(q,t;p) \phi_n(\tau;p) \vert 0 \rangle\,, \label{keyell}
\end{split}
\end{equation}
with $\phi_n(\tau;p) = \phi (\tau_1, \ldots, \tau_n; p)$ being the elliptic generalization of $\phi_n(\tau)$. With $p$ turned on Hamiltonian $\mathcal{O}_1(q,t;p)$ and its companions $\mathcal{O}_r(q,t;p)$ have been proposed in \cite{Koroteev:2015dja} to map onto quantum Hamiltonians of $\Delta$ILW hydrodynamic system. This observation was made based on the results of \cite{2009JPhA...42N4018S} reagrding the classical system. The conjecture 
\begin{equation}
\lim_{n \rightarrow \infty} \left[ t^{-n} \prod_{i=1}^n \dfrac{\Theta_p(qt^{-1}z/\tau_i)}{\Theta_p(qz/\tau_i)} \dfrac{\Theta_p(tz/\tau_i)}{\Theta_p(z/\tau_i)} \eta(z;pq^{-1}t) \right]_1  \vert 0 \rangle = 0 \label{extra}
\end{equation} 
of \cite{Feigin:2009ab} reduces at the level of eigenvalues to
\begin{equation}
\mathcal{E}_1^{(\lambda)}(p) = \lim_{n \rightarrow \infty} \left[ t^{-n+1}(1-t^{-1})\dfrac{(pt^{-1};p)_{\infty}(ptq^{-1};p)_{\infty}}{(p;p)_{\infty}(pq^{-1};p)_{\infty}} E_{eRS}^{(\lambda;n)}(p) \right]\,, \label{elllim}
\end{equation}
where $\mathcal{E}_1^{(\lambda)}(p)$ is the eigenvalue of $\mathcal{O}_1(q,t;p)$. 

\subsection{Bethe Ansatz Equations for $\Delta\text{ILW}$ from the ADHM Theory} \label{Sec:sec2.3}
In order to verify \eqref{elllim} one needs to know both $E_{eRS}^{(\lambda;n)}(p)$ for generic $n$ and $\mathcal{E}_1^{(\lambda)}(p)$. We already know from \eqref{WSU} that $E_{eRS}^{(\lambda;n)}(p)$ can be computed from the gauge theory, in particular we have
\begin{equation}
\dfrac{(pt^{-1};p)_{\infty}(ptq^{-1};p)_{\infty}}{(p;p)_{\infty}(pq^{-1};p)_{\infty}} E_{eRS}^{(\lambda;n)}(p) 
= \left\langle W_{\square}^{U(1)} \right\rangle E_{eRS}^{(\lambda;n)}(p) = \left\langle W_{\square}^{U(n)} \right\rangle \Big\vert_{\lambda}\,,
\end{equation}
where we identify $Q = e^{-8\pi^2 \gamma / g_{YM}^2}$ with $p$. What about $\mathcal{E}_1^{(\lambda)}(p)$? There are two ways of obtaining this eigenvalue:
\begin{itemize}
\item The most immediate possibility is to look for eigenstates of $\mathcal{O}_1(q,t;p)$ of the form $\sum_i c_i a_{-n_i}^{l_i}$ with fixed eigenvalue $k$ of the number operator $\sum_{n \geqslant 1}^{\infty} a_{-n}a_n$. We shall often refer to integer $k$ as the \textit{soliton number}. These states are in one-to-one correspondence with partitions of $k$. This method has the advantage that provides both eigenvalues and eigenfunctions and gives results exact in $p$, however, it becomes quickly computationally cumbersome for large $k$.

\item Alternatively we can use supersymmetric gauge theories again. As it was proposed in \cite{Ntalk,Otalk,NOinp} and further explored in \cite{2013JHEP...11..155L,2014JHEP...07..141B,2015JHEP...02..150A,2015arXiv150507116B}, the Coulomb branch of the Abelian (i.e. $N=1$) 2d ADHM gauge theory with gauge group $U(1)$ and a flavor group $U(N)$ is related via Bethe/Gauge correspondence \cite{Nekrasov:2009uh,Nekrasov:2009ui} to the ILW hydrodynamic system. Based on this observation, in \cite{Koroteev:2015dja} we proposed that the $\Delta$ILW system maps onto the 3d Abelian ADHM theory. In this setting eigenvalue $\mathcal{E}_1^{(\lambda)}(p)$ is given by\footnote{Here we are setting the parameter $a_1$ of Appendix \ref{appA} to zero.}
\begin{equation}
\mathcal{E}_1^{(\lambda)} = 1 - (1-q)(1-t^{-1}) \left\langle \text{Tr}\,\sigma \right\rangle \Big\vert_{\lambda}\,, \label{chern}
\end{equation}
i.e by the equivariant Chern character of the universal $U(1)$ bundle over the instanton moduli space. The local observable $\left\langle \text{Tr}\,\sigma \right\rangle$ with $\text{Tr}\,\sigma = \sum_s \sigma_s$ is evaluated at a solution $\lambda$ of the Bethe Ansatz equations \eqref{BAE2app} (solutions to these equations are once again labelled by partitions $\lambda$). 
We refer the reader to Appendix \ref{appA} for details on the ADHM theory.
\end{itemize}
Eigenvalue $\mathcal{E}_1^{(\lambda)}$ was computed using both methods in \cite{Koroteev:2015dja} for all possible partitions up to $k=3$, perturbatively in $p$, and the results of the computation have been shown to agree. This provides further evidence to the proposal of considering 3d Abelian ADHM as the gauge theory describing $\Delta$ILW system.

\subsection{$\Delta\text{ILW}$ as Large $n$ Limit of the Elliptic Ruijsenaars-Schneider Model} \label{Sec:sec2.4}
Having obtained both $E_{eRS}^{(\lambda;n)}(p)$ and $\mathcal{E}_1^{(\lambda)}(p)$ from computations in supersymmetric gauge theories, we can now check the validity of equation \eqref{elllim} which in gauge theoretic terms becomes
\begin{equation}
1 - (1-q)(1-t^{-1}) \left\langle \text{Tr}\,\sigma \right\rangle \Big\vert_{\lambda} =
\lim_{n \rightarrow \infty} \left[ t^{-n+1}(1-t^{-1}) \left\langle W_{\square}^{U(n)} \right\rangle\Big\vert_{\lambda} \right] \,. \label{gaugeenlim}
\end{equation}
This was the main computation carried out in \cite{Koroteev:2015dja} which illustrates a sophisticated relation between the large-$n$ asymptotics of the 5d $\mathcal{N}=1^*$ theory and the 3d $\mathcal{N}=2^*$ Abelian ADHM theory. 

Remarkably the above correspondence can be independently formulated using languages of three areas of mathematical physics: supersymmetric gauge theories, geometric representation theory and integrable many-body systems.

\begin{itemize}
\item
Let us first make a physics summary. In a given topological sector $k=\sum_i\lambda_i$ the left hand side of \eqref{gaugeenlim} computes onshell values of the Coulomb branch scalar of the 3d ADHM quiver theory describing moduli space of $k$ $U(1)$ instantons $\mathcal{M}_{k,1}$. Whereas the right hand side computes the large-$n$ regime of the fundamental $U(n)$ Wilson loop of the 5d theory evaluated at a locus \eqref{locus} of its Higgs branch. Morally speaking, $U(1)\subset U(n)$ factor survives through the large-$n$ transition and forms a gauge group of a different gauge theory. This explains why we identified FI parameter of the ADHM theory $p$ with the 5d instanton parameter $Q$.
\item
Mathematically we claim that there exists a stable limit of the equivariant Chern character of the universal bundle over the $U(n)$ instanton moduli space in terms of the same character only for $\mathcal{M}_{k,1}$. Other mathematical implications are listed in \cite{Koroteev:2015dja}. 
\item
Finally, from the point of view of integrable systems \eqref{gaugeenlim} states that the quantum spectrum of the elliptic RS model restricted on \eqref{locus} in the limit when the number of its particles becomes large is in one-to-one correspondence with the spectrum of twisted $\widehat{A}_0$ spin chain with one site and $k$ excitations. The twist parameter is given by elliptic deformation parameter $p$ of the eRS model. In addition the spin chain describes the $k$-soliton spectrum of the $\Delta$ILW system.
\end{itemize}

\subsection{Non-Abelian Generalization} \label{Sec:sec2.5}
Let us lastly summarize the main points of \cite{Koroteev:2015dja} which we have reviewed in this Section.  
\begin{enumerate}
\item We start by considering the $n$-particle trigonometric and elliptic Ruijsenaars models. As was analyzed in \cite{Bullimore:2015fr} their eigenfunctions and eigenvalues can be obtained from computations of BPS observables in the 5d $\mathcal{N}=1^*$ $U(n)$ theory in the presence of defects (see \eqref{WSU}).
\item We then realize tRS and eRS in terms of free fields along the lines of \cite{Feigin:2009ab}. In the limit of the large number of particles this realization provides an infinite number of commuting quantum Hamiltonians, which we identify with the Hamiltonians of the $\Delta$ILW system. This suggests large-$n$ relation \eqref{elllim} between the eRS and the $\Delta$ILW spectra.
\item We compute the $\Delta$ILW eigenvalues in two ways: first directly from the Hamiltonian, and then as a local observable of the 3d Abelian ($N=1$) ADHM theory \eqref{chern}. We check that the two computations agree, which implies that the 3d Abelian ADHM theory is related to the $\Delta$ILW system.
\item Finally, knowing both eRS and $\Delta$ILW eigenvalues we verify proposal \eqref{elllim} for first several topological sectors. This proposal, being somewhat intuitive from the integrable system point of view, yields the non-trivial relation \eqref{gaugeenlim} between a Wilson loop in the 5d $\mathcal{N}=1^*$ $U(n)$ theory and a local observable in the 3d Abelian ADHM theory.
\end{enumerate}

In this paper we will address the following question: \textit{What happens if we consider the natural non-Abelian ($N > 1$) generalization of the 3d ADHM theory? Which integrable system and which gauge theory in the large-$n$ will lead us to this model?}

It turns out that one can naturally realize a system of $N$ coupled eRS systems ($\text{eRS}_N$) in terms of free fields (although we are not aware of an expression for the Hamiltonians of this system in terms of finite-difference operators like \eqref{eRSh1}\footnote{The closest analogue that we are aware of is given in Proposition A.10 of \cite{Feigin:2009ab}.}) again in the setting of \cite{Feigin:2009ab} (see Appendix A there). This is related to the level-$N$ representation of the Ding-Iohara algebra.

At the trigonometric level, we will obtain a system of $N$ coupled tRS systems ($\text{tRS}_N$) written in terms of free fields. Their eigenfunctions are called \textit{generalized Macdonald polynomials} in the literature (see for example \cite{2011arXiv1106.4088A,Belavin:2012eg,Zenkevich:2014lca}). Although we will not comment further on this, the $\text{tRS}_N$ systems have deep connections to the 5d analogue of the AGT conjecture \cite{Alday:2009aq,Wyllard:2009hg}. Similar $N$ coupled trigonometric Calogero-Sutherland systems and generalized Jack polynomials have appeared in \cite{2011LMaPh..98...33A,2012JHEP...01..051F,2012NuPhB.860..377E,2011NuPhB.850..199B} in relation with the original 4d AGT correspondence: in \textit{loc cit} it was noticed that the Hamiltonian possesses an interesting triangular structure (which appears also in the $\text{tRS}_N$ case), which implies that the eigenvalues of the system are simply given by the sum of the eigenvalues of $N$ decoupled trigonometric Calogero-Sutherland models.\footnote{This is equivalent to say that the eigenvalues of an upper-triangular matrix do not depend on the non-diagonal entries.} Moreover, the infinite set of commuting quantum Hamiltonians arising from the free field realization of these trigonometric $N$ coupled Calogero models have been related to the Benjamin-Ono limit of the so-called $gl(N)$ Intermediate Long Wave hydrodynamic systems \cite{lebedev1983,ref1,1992JMP....33.3783D}. On the other hand, the $gl(N)$ ILW has been related to the elliptic coupled $N$ copies of Calogero models in \cite{Ntalk,Otalk,NOinp,2013JHEP...11..155L,2014JHEP...07..141B,2015JHEP...02..150A,2015arXiv150507116B}, as well as to the quantum cohomology of the instanton moduli space (\cite{Ntalk,Otalk,NOinp,2014JHEP...07..141B,2015arXiv150507116B} and especially \cite{2012arXiv1211.1287M}). 

Keeping this in mind, we propose that the $n$-particle $\text{eRS}_N$ model will reduce at large $n$ to what we would call the quantum $gl(N)$ $\Delta\text{ILW}$ system (or $\Delta\text{ILW}_N$). We are not aware of any study on a similar system in the hydrodynamic literature, apart from the already mentioned $\Delta\text{ILW}_1 = \Delta\text{ILW}$ classical case \cite{2009JPhA...42N4018S} and its $\Delta\text{BO}_1 = \Delta\text{BO}$ limit \cite{2009arXiv0911.5005T,2011ntqi.conf..357S}. In any case, we can define quantum $\Delta\text{ILW}_N$ as the system corresponding to the infinite number of commuting quantum Hamiltonians which arise from the free field construction of $\text{eRS}_N$, in the limit of infinite number of particles. Then we will proceed with the strategy from the list above, albeit in a slightly different order:
\begin{enumerate}
\item[2.] We start in Section \ref{Sec:sec3} by defining $\text{tRS}_N$ and $\text{eRS}_N$ in terms of free fields as in Appendix A of \cite{Feigin:2009ab}. In the large number of particles limit this realization provides an infinite number of commuting quantum Hamiltonians, which we identify with the $\Delta\text{ILW}_N$ operators, and suggests a large-$n$ relation between the $\text{eRS}_N$ and $\Delta\text{ILW}_N$ spectra (as well as between their $\text{tRS}_N$ and $\Delta\text{BO}_N$ cousins).
\item[3.] We compute the $\Delta\text{ILW}_N$ eigenvalues in two ways: first directly from the Hamiltonian (Section \ref{Sec:sec3}), and then as a local observable in the 3d non-Abelian ($N>1$) ADHM theory (Section \ref{Sec:sec4}). We check that the two computations agree, which suggests that the 3d non-Abelian ADHM theory is dual to the $\Delta\text{ILW}_N$ system.
\item[1.] Based on the analogy with the $N=1$ model we propose that the $Nn$-particles $\text{eRS}_N$ eigenvalue can be computed from the gauge theory as the vacuum expectation value of the Wilson loop in the fundamental representation of the 5d $\mathcal{N}=1^*$ $U(Nn)$ theory (Section \ref{Sec:sec5}).
\item[4.] Finally in \Secref{Sec:sec5} we verify that the $\text{eRS}_N$ model reduces to the $\Delta\text{ILW}_N$ system in the sense which we described above. Our proposal implies an equality between the Wilson loop in the 5d $\mathcal{N}=1^*$ $U(Nn)$ theory and a Coulomb branch scalar in the 3d non-Abelian ADHM theory.
\end{enumerate}
The rest of the paper explains in greater details the above points.

\section{Free Field Realization of $N$ Coupled Ruijsenaars-Schneider Systems and $\Delta\text{ILW}_N$ Models} \label{Sec:sec3}
Let us start by constructing the eRS$_N$ system in the free field formalism following the procedure described in appendix A of \cite{Feigin:2009ab}. The basic ingredients were presented in \secref{Sec:sec2.2}. We consider $N$ copies of the $(q,t)$-deformed Heisenberg algebra $\oplus_{l=1}^N \mathcal{H}^{(l)}(q,t)$, generated by the modes $a_m^{(l)}$, $m \in \mathbb{Z}$, $l = 1, \ldots, N$ following the commutation relation
\begin{equation}
[a_m^{(l)},a_n^{(r)}] = m \dfrac{1-q^{\vert m \vert}}{1-t^{\vert m \vert}} \delta^{l,r} \delta_{m+n,0} \,.
\end{equation}
Next we introduce elliptic vertex operator $\eta^{(l)}(z)$:
\begin{equation}
\eta^{(l)}(z;p^{(l)}) = \text{exp} \left( \sum_{n>0} \dfrac{1-t^{-n}}{n} \dfrac{1-(p^{(l)})^n}{1-(p^{(l)} q t^{-1})^n} a_{-n}^{(l)} z^n \right) \text{exp} \left( -\sum_{n>0} \dfrac{1-t^n}{n} a_n^{(l)} z^{-n} \right)\,, \label{etapN}
\end{equation}
where, as we will see shortly, parameters $p^{(l)}$ are proportional to the parameter of elliptic deformation $p$.\footnote{Notice the shift which we performed on the $p^{(l)}$ in \eqref{etapN} compared to the previous definition \eqref{etap}.}
Similarly, we introduce vertex operators
\begin{equation}
\begin{split}
\varphi^{(l)}_-(z;p^{(l)}) \;=\; & \text{exp} \left( \sum_{n>0} \dfrac{1-t^{-n}}{n} (1-(q^{-1}t)^n) \dfrac{1}{1-(p^{(l)}q t^{-1})^n}(qt^{-1})^{\frac{n}{4}} a_{-n}^{(l)} z^n \right) \\
& \text{exp} \left( -\sum_{n>0} \dfrac{1-t^n}{n} (1-(q^{-1}t)^n) \dfrac{(p^{(l)}q t^{-1})^n}{1-(p^{(l)})^n}(qt^{-1})^{-\frac{n}{4}} a_n^{(l)} z^{-n} \right)\,. \label{phi-pN}
\end{split}
\end{equation}
Although these operators do not appear in the $N=1$ construction, they emerge naturally in the present formalism, since they are part of the realization of the underlying Ding-Iohara algebra \cite{Feigin:2009ab}. 

From \eqref{phi-pN} one can construct an infinite family of quantum commuting operators, which we identify as the ones of the $\Delta\text{ILW}_N$ system, in particular the first Hamiltonian $\widehat{\mathcal{H}}_1^{(N)}$ is given by
\begin{equation}
\widehat{\mathcal{H}}_1^{(N)} = \left[ \sum_{l=1}^N \alpha_l \widetilde{\Lambda}_l^{(N)}(z) \right]_1 = 
\dfrac{1}{2\pi i} \oint \dfrac{dz}{z} \left( \sum_{l=1}^N \alpha_l \widetilde{\Lambda}_l^{(N)}(z) \right)\,, \label{ham}
\end{equation}
where
\begin{equation}
\widetilde{\Lambda}_l^{(N)}(z) = \left( \prod_{r=1}^{l-1} \varphi_-^{(r)}\left((q^{-1}t)^{\frac{2r-1}{4}}z; p(q t^{-1})^{N-r}\right) \right) \eta^{(l)}\left((q^{-1}t)^{\frac{l-1}{2}}z; p(q t^{-1})^{N-l}\right)\,, \label{lambda}
\end{equation}
and $\alpha_l$ are a set of complex parameters. When $N=1$ Hamiltonian \eqref{ham} reduces to  
\begin{equation}
\widehat{\mathcal{H}}_1^{(1)} = \left[ \alpha_1 \eta\left(z; p\right) \right]_1 
\end{equation}
which (for $\alpha_1 = 1$) corresponds to operator $\mathcal{O}_1(q,t;p)$ which we have used in \secref{Sec:sec2.2},  whose eigenvalues were studied in detail in \cite{Koroteev:2015dja}. 

In the trigonometric limit we expect the Hamiltonian \eqref{ham} to be related to the first $\text{tRS}_N$ Hamiltonian in a way similar to \eqref{keytr} and in particular
\begin{equation}
\widehat{\mathcal{H}}_{1;\Delta\text{BO}_N}^{(N)} \vert \psi \rangle = \left[ t^{-n} \sum_{l=1}^N \alpha_l + t^{-n+1}(1-t^{-1})D^{(1;N)}_{Nn,\vec{\tau};\text{tRS}_N}(q,t) \right] \vert \psi \rangle \label{keytrN} \,.
\end{equation}
At the level of eigenvalues this would imply a relation similar to \eqref{trigeig}, which in this case reads\footnote{Notice the power $n$ in the exponents of $t$ instead of $nN$. This choice is justified by the computations in Section \ref{Sec:sec5}.}
\begin{equation}
\mathcal{E}^{(N;\vec{\lambda})}_{1;\Delta\text{BO}_N} = t^{-n} \sum_{l=1}^N \alpha_l + t^{-n+1}(1-t^{-1})E_{\text{tRS}_N}^{(Nn;\vec{\lambda})} \label{questa}
\end{equation}
which in the $n \rightarrow \infty$ limit and for $\vert t \vert < 1$ becomes
\begin{equation}
\mathcal{E}_{1;\Delta\text{BO}_N}^{(N;\vec{\lambda})} = \lim_{n \rightarrow \infty} \left[ t^{-n+1}(1-t^{-1})E_{\text{tRS}_N}^{(Nn;\vec{\lambda})} \right] \,.
\end{equation}  
In an analogous way, we expect the Hamiltonian \eqref{ham} to be related to the first $Nn$-particles $\text{eRS}_N$ Hamiltonian in a way similar to \eqref{keyell}; we anticipate here that at the level of eigenvalues this relation leads in the $n \rightarrow \infty$ limit to
\begin{equation}
\mathcal{E}_{1;\Delta\text{ILW}_N}^{(N;\vec{\lambda})}(p) = \lim_{n \rightarrow \infty} \left[ t^{-n+1}(1-t^{-1})\dfrac{(Qt^{-1};Q)_{\infty}(Qtq^{-1};Q)_{\infty}}{(Q;Q)_{\infty}(Qq^{-1};Q)_{\infty}} E_{\text{eRS}_N}^{(Nn;\vec{\lambda})}(p) \right]\,, \label{elllimN}
\end{equation}
where $Q$ is proportional to $p$. The above equality will be thoroughly verified in Section \ref{Sec:sec5}. Clearly in order to do this check we will need to know both the eigenvalue $\mathcal{E}^{(N)}_1$ of the first $\Delta\text{ILW}_N$ Hamiltonian \eqref{ham} and the eigenvalue $E_{1}^{(Nn)}$ of the first $\text{eRS}_N$ Hamiltonian for generic $n$. In Section \ref{Sec:sec5} we will compute the latter; the former will be computed in the remaining part of this Section directly by considering eigenstates of $\widehat{\mathcal{H}}_1^{(N)}$, and in Section \ref{Sec:sec4} indirectly by considering local observables in the 3d ADHM non-Abelian theory. \\

\noindent Some clarifications are needed at this point, since what we are doing is only well defined in the $N=1$ case. 
In fact, when $N=1$ we know the explicit expressions of both the tRS and eRS Hamiltonians in terms of finite-difference operators: these are given by \eqref{tRSh1} and \eqref{eRSh1} respectively. We now want to consider them in terms of free fields. In the trigonometric case we have \eqref{keytr}, where we defined $[\eta(z)]_1$ to be the first Hamiltonian of the quantum $\Delta\text{BO}$ system because its classical limit reduces to the classical $\Delta\text{BO}$ system of \cite{2011ntqi.conf..357S,2009arXiv0911.5005T}. Here $D^{(1)}_{n,\vec{\tau}}(q,t)$ and $[\eta(z)]_1$ can be thought to have the same eigenstates (Macdonald polynomials $P_{\lambda}$ and Macdonald eigenstates $\vert P_{\lambda} \rangle$, obtained from the polynomials by making use of the isomorphism \eqref{isomorphism}) but they have different eigenvalues, both computable. Similarly, in the elliptic case we have \eqref{keyell}, and we defined $[\eta(z;pqt^{-1})]_1$ as the first Hamiltonian of the quantum $\Delta\text{ILW}$ system because in the classical limit it reduces to the classical $\Delta\text{ILW}$ system studied in \cite{2009JPhA...42N4018S}. Analogously $D^{(1)}_{n,\vec{\tau}}(q,t;p)$ and $[\eta(z;pqt^{-1})]_1$ do not have the same eigenstates\footnote{There should be a precise sense for which the two have the same eigenstates at large $n$, but this is not yet clear to us at the moment.}, the isomorphism \eqref{isomorphism} can no longer be applied, but we can nevertheless compute both eigenvalues (thanks to gauge theory complutations) and check the validity of the large-$n$ relation \eqref{elllim} between the two spectra. 

On the other hand, when $N>1$ we do not have explicit expressions for the $\text{tRS}_N$ and $\text{eRS}_N$ Hamiltonians in terms of finite-difference operators, nor do we have classical limits of the $\Delta\text{BO}_N$ and $\Delta\text{ILW}_N$ Hamiltonians. Instead we \textit{define} them in a way which naturally generalizes the $N=1$ construction and is which is consistent with the gauge theory results. For example, we define the $\Delta\text{BO}_N$ and $\Delta\text{ILW}_N$ Hamiltonians as in \eqref{ham} since operator \eqref{ham} is the natural extension of $[\eta(z;pqt^{-1})]_1$ when one considers level $N$ representations of the Ding-Iohara algebra underlying our free field construction, and also because (as we will see in the next Section) its eigenvalue coincides with the observable of the non-Abelian ADHM theory that naturally generalizes \eqref{chern}. Although we cannot be completely certain that our physics-motivated construction describes proper finite-difference hydrodynamic systems, we nevertheless have enough grounds to believe in the validity of our proposal based on what is know about the differential $gl(N)$ ILW and $gl(N)$ BO systems. 

The $\text{tRS}_N$ instead is defined by \eqref{keytrN}, based on the analogy with the $N=1$ case: it is simply a shift of the $\Delta\text{BO}_N$ Hamiltonian by a constant term.\footnote{Our choice for the $\text{tRS}_N$ Hamiltonian is different from the ones made in \cite{2011arXiv1106.4088A,Zenkevich:2014lca}; in particular \cite{2011arXiv1106.4088A} defines the $\text{tRS}_N$ Hamiltonian as our $\Delta\text{BO}_N$ Hamiltonian, while the one in \cite{Zenkevich:2014lca} corresponds to our $\Delta\text{BO}_N$ Hamiltonian with the additional term $-\sum_{l=1}^N \alpha_l$. All of those operators have the same set of eigenstates but different eigenvalues; in the choice of \cite{Zenkevich:2014lca} the vacuum state has zero energy. We think the name $\text{tRS}_N$ is more appropriate for our choice, since in the $N=1$ case it reduces to the original trigonometric Ruijsenaars-Schneider Hamiltonian \eqref{tRSh1} without any additional constant term.} Again, the $\Delta\text{BO}_N$ and $\text{tRS}_N$ Hamiltonians will have the same eigenstates (in the opposite order: generalized Macdonald eigenstates $\vert P_{\vec{\lambda}} \rangle$ and generalized Macdonald polynomials $P_{\vec{\lambda}}$, obtained from the eigenstates by making use of the isomorphism \eqref{isomorphism}) but they have different eigenvalues, both computable. As we will later see in Section \ref{Sec:sec5} the eigenvalues of $\text{tRS}_N$ have a natural gauge theory interpretation as a Wilson loop in the 5d $U(Nn)$ $\mathcal{N}=1^*$ theory with 5d gauge coupling turned off, which is an immediate generalization of what we had in the $N=1$ case. 

Finally, as far as $\text{eRS}_N$ is concerned we shall assume that the eigenvalues of its Hamiltonian are given by the fundamental Wilson loop in the 5d $U(Nn)$ $\mathcal{N}=1^*$ theory with 5d gauge coupling turned on, and that relations analogous to \eqref{keyell}, \eqref{extra} also take place. Had we have an explicit construction for the $\text{tRS}_N$ Hamiltonian in terms of finite-difference operators, we would expected the $\text{eRS}_N$ Hamiltonian to be its natural elliptization. In the following however we will only be concerned with its eigenvalues, which as will show in Section \ref{Sec:sec5} obey \eqref{elllimN}.\\

\noindent  Having clarified the above subtleties, let us proceed to the computation of $\mathcal{E}^{(N)}_1$ from \eqref{ham}. We first rewrite \eqref{etapN} and \eqref{phi-pN} as
\begin{equation}
\eta^{(l)}(z;p^{(l)}) = \text{exp}\left(\sum_{n>0}\lambda_{-n}^{(l)} z^n\right) \text{exp}\left(\sum_{n>0}\lambda_{n}^{(l)} z^{-n}\right)\,, 
\end{equation}
\begin{equation}
\varphi^{(l)}_-(z;p^{(l)}) = \text{exp}\left(\sum_{n>0}\omega_{-n}^{(l)}z^n\right) \text{exp}\left(\sum_{n>0}\omega_{n}^{(l)}z^{-n}\right)\,, 
\end{equation}
with commutation relations 
\begin{equation}
[\lambda_m^{(l)}, \lambda_n^{(r)}] = -\dfrac{1}{m} \dfrac{(1-q^m)(1-t^{-m})(1-(p^{(l)})^m)}{1-(p^{(l)}q t^{-1})^m} \delta_{m+n,0} \; \delta^{l,r}
\end{equation}
for the $\lambda_m^{(l)}$ and 
\begin{equation}
[\omega_m^{(l)}, \omega_n^{(r)}] = -\dfrac{1}{m} \dfrac{(1-q^m)(1-t^{-m})(1-(q^{-1}t)^m)^2}{(1-(p^{(l)}q t^{-1})^m)(1-(p^{(l)})^m)} (p^{(l)}q t^{-1})^m \; \delta_{m+n,0} \; \delta^{l,r} 
\end{equation}
for the $\omega_m^{(l)}$.
By comparing \eqref{etapN} with \eqref{phi-pN} we conclude that
\begin{equation}
\begin{split}
\omega_{-m}^{(l)} & = \dfrac{1-(q^{-1}t)^m}{1-(p^{(l)})^m}(qt^{-1})^{\frac{m}{4}} \lambda_{-m}^{(l)} \\
\omega_{m}^{(l)} & = \dfrac{1-(q^{-1}t)^m}{1-(p^{(l)})^m}(qt^{-1})^{\frac{3m}{4}} (p^{(l)})^{m} \lambda_{m}^{(l)}
\end{split}
\end{equation}
Notice here that $\omega_m^{(l)} \rightarrow 0$ ($m>0$) in the $\Delta\text{BO}_N$ (trigonometric) limit $p \rightarrow 0$, although $\omega_{-m}^{(l)}$ does not vanish: we therefore explicitly see the triangular structure of the $\Delta\text{BO}_N$ (and $\text{tRS}_N$) system we mentioned in Section \ref{Sec:sec2.5}. As a consequence of this structure, the spectrum at $p \rightarrow 0$ is merely given by the sum of $N$ copies of the spectrum of $\Delta \text{BO}_{1}$ system (and similarly for $\text{tRS}_N$). 

The computation of $\mathcal{E}^{(N)}_1$ now proceeds as follows. At fixed $N$, we consider linear combination of the annihilation operators $\lambda_{-m}^{(l)}$ of level $k$ (the eigenvalue of the number operator $\sum_{l=1}^N\sum_{m>0}a_{-m}^{(l)}a_m^{(l)}$) acting on the vacuum $\vert 0 \rangle$. The number of possible coefficients of the linear combination naturally coincides with the number of $N$-partitions $\vec{\lambda} = (\lambda^{(1)}; \ldots; \lambda^{(N)})$ of $k$. Requiring this state with generic coefficients to be an eigenstate of \eqref{ham} will fix the energy eigenvalue and the coefficients themselves (modulo an overall normalization). Although this procedure provides expressions which are exact in $p$, we will truncate the solution to a low order in the small $p$ expansion. The computations presented here regard very low values of $k$ and $N$; higher values can certainly be considered straightforwardly, unfortunately it leads to the rapid increase of the possible eigenstates which dramatically increases the computational time.

\subsection{$\Delta\text{ILW}_2$ Spectrum} \label{Sec:sec3.1}
For the sake of clarity let us consider $N=2$. The first Hamiltonian reads
\begin{equation}
\widehat{\mathcal{H}}_1^{(2)} = \left[ \alpha_1 \eta^{(1)}\left(z; p^{(1)}\right) + \alpha_2  \varphi_-^{(1)}\left((q^{-1}t)^{\frac{1}{4}}z; p^{(1)}\right) \eta^{(2)}\left((q^{-1}t)^{\frac{1}{2}}z; p^{(2)}\right) \right]_1 
\end{equation}
where $p^{(1)} = p qt^{-1}$ and $p^{(2)} = p$. We will study eigenstates of $\widehat{\mathcal{H}}_1^{(2)}$ at fixed low $k$ to which we shall refer to as the soliton number.
 
\subsubsection{Zero solitons} \label{oscN2k0}
A generic state with $k=0$ is given by 
\begin{equation}
c_1 \vert 0 \rangle\,,
\end{equation}
therefore there is only one possible state -- the vacuum corresponding to the only 2-partition $(\bullet; \bullet)$ of $k=0$ modulo normalization. Acting on it with $\widehat{\mathcal{H}}_1^{(2)}$ we get
\begin{equation}
\widehat{\mathcal{H}}_1^{(2)} c_1 \vert 0 \rangle = \left[\alpha_1 + \alpha_2 \right] c_1 \vert 0 \rangle = \mathcal{E}_1^{(2)} c_1 \vert 0 \rangle
\end{equation}
therefore of our vacuum state is an eigenstate with eigenvalue
\begin{equation}
\mathcal{E}_1^{(2;(\bullet; \bullet))} = \alpha_1 + \alpha_2 .
\end{equation}

\subsubsection{One soliton} \label{oscN2k1}
A generic state with $k=1$ is given by 
\begin{equation}
(c_1 \lambda_{-1}^{(1)} + c_2 \lambda_{-1}^{(2)}) \vert 0 \rangle \,.
\end{equation}
In this case the eigenstate equation 
\begin{equation}
\begin{split}
& \widehat{\mathcal{H}}_1^{(2)} (c_1 \lambda_{-1}^{(1)} + c_2 \lambda_{-1}^{(2)}) \vert 0 \rangle = \mathcal{E}_1^{(2)} (c_1 \lambda_{-1}^{(1)} + c_2 \lambda_{-1}^{(2)}) \vert 0 \rangle = \\
& = \Big[ \alpha_1 + \alpha_2 + \alpha_1 \lambda_{-1}^{(1)}\lambda_{1}^{(1)} + \alpha_2 \lambda_{-1}^{(2)} \lambda_{-1}^{(2)} 
+ \alpha_2 p q^2 t^{-2}\left( \dfrac{1-q^{-1}t}{1-pq t^{-1}} \right)^2 \lambda_{-1}^{(1)}\lambda_{1}^{(1)} \\
& + \alpha_2 (qt^{-1})^{\frac{1}{2}} \dfrac{1-q^{-1}t}{1-pq t^{-1}} \lambda_{-1}^{(1)}\lambda_{1}^{(2)}
+ \alpha_2 (qt^{-1})^{\frac{3}{2}} p \dfrac{1-q^{-1}t}{1-pq t^{-1}} \lambda_{-1}^{(2)}\lambda_{1}^{(1)} \Big](c_1 \lambda_{-1}^{(1)} + c_2 \lambda_{-1}^{(2)}) \vert 0 \rangle
\end{split}
\end{equation}
%\begin{equation}
%\text{Det} 
%\left( \begin{array}{cc}
%\alpha_1 + \alpha_2 - \alpha_1 \dfrac{(1-q)(1-t^{-1})(1-p)}{1-pqt^{-1}} - \alpha_2 \dfrac{(1-q)(1-t^{-1})(1-p)}{1-pqt^{-1}} \left(\dfrac{1-q^{-1}t}{1-p}\right)^2 pqt^{-1} - \mathcal{E}_1 & - \alpha_2 \dfrac{1-q^{-1}t}{1-p}(pqt^{-1})^{\frac{1}{2}}\dfrac{(1-q)(1-t^{-1})(1-pq^{-1}t)}{1-p}(qt^{-1})^{-\frac{1}{4}} \\ 
%- \alpha_2 \dfrac{1-q^{-1}t}{1-p}(pqt^{-1})^{\frac{1}{2}}\dfrac{(1-q)(1-t^{-1})(1-p)}{1-pqt^{-1}}(qt^{-1})^{\frac{1}{4}} & \alpha_1 + \alpha_2 - \alpha_2 \dfrac{(1-q)(1-t^{-1})(1-pq^{-1}t)}{1-p} - \mathcal{E}_1
%\end{array} \right) = 0 
%\end{equation}
admits two possible solutions, which can be labelled by two 2-partitions $(\square;\bullet)$ and $(\bullet;\square)$ of $k=1$; the corresponding eigenvalues are 
\begin{equation}
\begin{split}
\mathcal{E}_1^{(2;(\square; \bullet))} & \;=\; \alpha_1 (q + t^{-1} - qt^{-1}) + \alpha_2 + p\, \alpha_1 \dfrac{(1-q)(1-t)(q-t)(q\alpha_1 - t\alpha_2)}{t^3(\alpha_1 - \alpha_2)} \\
& + p^2 \alpha_1 \frac{(1-q) (1-t) (q-t) \left(\alpha_1 q - \alpha_2 t\right) \left(\alpha_1^2 q^2-\alpha_1 \alpha_2 \left(2 q^2-q t+t^2\right)+\alpha_2^2 q t\right)}{t^5\left(\alpha_1 -\alpha_2\right)^3} + o(p^3) \\ 
\mathcal{E}_1^{(2;(\bullet; \square))} & \;=\; \alpha_1 + \alpha_2 (q + t^{-1} - qt^{-1}) + p\, \alpha_2 \dfrac{(1-q)(1-t)(q-t)(q\alpha_2 - t\alpha_1)}{t^3(\alpha_2 - \alpha_1)} \\
& + p^2 \alpha_2 \frac{(1-q) (1-t) (q-t) \left(\alpha_2 q - \alpha_1 t\right) \left(\alpha_2^2 q^2-\alpha_1 \alpha_2 \left(2 q^2-q t+t^2\right)+\alpha_1^2 q t\right)}{t^5\left(\alpha_2 -\alpha_1\right)^3} + o(p^3) \,.
\end{split}
\end{equation}
As a side comment, let us remark that for $p=0$ the eigenstates are given by
\begin{equation}
\begin{split}
& \vert \square; \bullet \rangle = c_1 \lambda^{(1)}_{-1} \vert 0 \rangle \\
& \vert \bullet; \square \rangle = c_2 \left[ \sqrt{qt^{-1}} \dfrac{\alpha_2(t-q)}{q(\alpha_1 - \alpha_2)} \lambda^{(1)}_{-1} + \lambda^{(2)}_{-1} \right] \vert 0 \rangle
\end{split}
\end{equation}
and correspond to the $N=2$, level $k=1$ generalized Macdonald polynomials given in \cite{2011arXiv1106.4088A,Belavin:2012eg,Zenkevich:2014lca}. Although not explicitly written here, similar results can also be obtained for the other cases considered in the following.

\subsubsection{Two solitons} \label{oscN2k2}

A generic state with $k=2$ is given by 
\begin{equation}
\left[ c_1 \left( \lambda_{-1}^{(1)} \right)^2 + c_2 \lambda_{-2}^{(1)} + c_3 \left( \lambda_{-1}^{(2)} \right)^2 + c_4 \lambda_{-2}^{(2)} + c_5 \lambda_{-1}^{(1)} \lambda_{-1}^{(2)} \right] \vert 0 \rangle
\end{equation} 
The eigenstate equation
\begin{equation}
\begin{split}
& \widehat{\mathcal{H}}_1^{(4)} (c_1 \left( \lambda_{-1}^{(1)} \right)^2 + c_2 \lambda_{-2}^{(1)} + c_3 \left( \lambda_{-1}^{(2)} \right)^2 + c_4 \lambda_{-2}^{(2)} + c_5 \lambda_{-1}^{(1)} \lambda_{-1}^{(2)}) \vert 0 \rangle \\
& = \mathcal{E}_1^{(2)} (c_1 \left( \lambda_{-1}^{(1)} \right)^2 + c_2 \lambda_{-2}^{(1)} + c_3 \left( \lambda_{-1}^{(2)} \right)^2 + c_4 \lambda_{-2}^{(2)} + c_5 \lambda_{-1}^{(1)} \lambda_{-1}^{(2)}) \vert 0 \rangle
\end{split}
\end{equation}
admits five solutions, labelled by the five 2-partitions $(\tiny{\yng(1,1)}; \bullet)$, $(\tiny{\yng(2)}; \bullet)$, $(\tiny{\yng(1,1)}; \bullet)$, $(\bullet; \tiny{\yng(2)})$, $(\tiny{\yng(1)}; \tiny{\yng(1)})$ of $k=2$; the corresponding eigenvalues are 
\begin{equation}
\begin{split}
\mathcal{E}^{(2;(\tiny{\yng(1,1)}; \bullet))}_1 & \;=\; \alpha_1(q^2 + t^{-1} - q^2t^{-1}) + \alpha_2 
+ p \, \alpha_1 \dfrac{q(1-q^2)(1-t)^2(q-t)(q^2 \alpha_1 - t \alpha_2)}{t^3(1-qt)(q\alpha_1 - \alpha_2)} + o(p^2) \\ 
\mathcal{E}^{(2;(\tiny{\yng(2)}; \bullet))}_1 & \;=\; \alpha_1(q + t^{-2} - qt^{-2}) + \alpha_2 
+ p \, \alpha_1 \dfrac{(1-q)^2(1-t^2)(q-t)(q \alpha_1 - t^2 \alpha_2)}{t^4(1-qt)(\alpha_1 - t\alpha_2)} + o(p^2) \\ 
\mathcal{E}^{(2;(\tiny{\yng(1,1)}; \bullet))}_1 & \;=\; \alpha_1 + \alpha_2(q^2 + t^{-1} - q^2t^{-1}) 
+ p \, \alpha_2 \dfrac{q(1-q^2)(1-t)^2(q-t)(q^2 \alpha_2 - t \alpha_1)}{t^3(1-qt)(q\alpha_2 - \alpha_1)} + o(p^2) \\ 
\mathcal{E}^{(2;(\bullet; \tiny{\yng(2)}))}_1 & \;=\; \alpha_1 + \alpha_2(q + t^{-2} - qt^{-2}) 
+ p \, \alpha_2 \dfrac{(1-q)^2(1-t^2)(q-t)(q \alpha_2 - t^2 \alpha_1)}{t^4(1-qt)(\alpha_2 - t\alpha_1)} + o(p^2) \\ 
\mathcal{E}^{(2; (\tiny{\yng(1)}; \tiny{\yng(1)}))}_1 & \;=\; (\alpha_1 + \alpha_2) (q + t^{-1} - qt^{-1}) \\
& + \tilde{p} \alpha_1 \dfrac{(1-q)(1-t)(q-t)(q\alpha_1 - \alpha_2)(t\alpha_1 - q\alpha_2)(\alpha_1 - t\alpha_2)}{t^3(\alpha_1 - \alpha_2)(t\alpha_1 - \alpha_2)(\alpha_1 - q\alpha_2)} \\
& + \tilde{p} \alpha_2 \dfrac{(1-q)(1-t)(q-t)(q\alpha_2 - \alpha_1)(t\alpha_2 - q\alpha_1)(\alpha_2 - t\alpha_1)}{t^3(\alpha_2 - \alpha_1)(t\alpha_2 - \alpha_1)(\alpha_2 - q\alpha_1)} + o(\tilde{p}^2) \,.
\end{split}
\end{equation}

\subsection{$\Delta\text{ILW}_3$ Spectrum} \label{Sec:sec3.2}

In the $N=3$ case the first Hamiltonian reads
\begin{equation} 
\begin{split}
\widehat{\mathcal{H}}^{(3)}_1 = \Big[ & \alpha_1 \eta^{(1)}\left(z; p^{(1)} \right)
+ \alpha_2 \varphi_-^{(1)}\left((q^{-1}t)^{\frac{1}{4}}z; p^{(1)}\right) 
\eta^{(2)}\left((q^{-1}t)^{\frac{1}{2}}z; p^{(2)}\right) \\
& + \alpha_3 \varphi_-^{(1)}\left((q^{-1}t)^{\frac{1}{4}}z; p^{(1)}\right)
\varphi_-^{(2)}\left((q^{-1}t)^{\frac{3}{4}}z; p^{(2)}\right)
\eta^{(3)}\left((q^{-1}t) z; p^{(3)}\right)
\Big]_1
\end{split}
\end{equation}
with $p^{(1)}=p(qt^{-1})^2$, $p^{(2)} = pqt^{-1}$ and $p^{(3)} = p$.

\subsubsection{Zero solitons} \label{oscN3k0}

The only possible state with $k=0$ is the vacuum 
\begin{equation}
c_1 \vert 0 \rangle
\end{equation}
which corresponds to the only 3-partition $(\bullet; \bullet; \bullet)$ of $k=0$.
The $\widehat{\mathcal{H}}_1^{(3)}$ action
\begin{equation}
\widehat{\mathcal{H}}_1^{(3)} c_1 \vert 0 \rangle = \left[\alpha_1 + \alpha_2 + \alpha_3 \right] c_1 \vert 0 \rangle = \mathcal{E}_1^{(3)} c_1 \vert 0 \rangle
\end{equation}
gives the vacuum energy
\begin{equation}
\mathcal{E}_1^{(3;(\bullet; \bullet; \bullet))} = \alpha_1 + \alpha_2 + \alpha_3 \,.
\end{equation}

\subsubsection{One soliton} \label{oscN3k1}

A generic state with $k=1$ can be written as
\begin{equation}
(c_1 \lambda_{-1}^{(1)} + c_2 \lambda_{-1}^{(2)} + c_3 \lambda_{-1}^{(3)}) \vert 0 \rangle \,.
\end{equation}
The eigenstate equation
\begin{equation}
\widehat{\mathcal{H}}_1^{(3)} (c_1 \lambda_{-1}^{(1)} + c_2 \lambda_{-1}^{(2)} + c_3 \lambda_{-1}^{(3)}) \vert 0 \rangle
= \mathcal{E}_1^{(3)} (c_1 \lambda_{-1}^{(1)} + c_2 \lambda_{-1}^{(2)} + c_3 \lambda_{-1}^{(3)}) \vert 0 \rangle
\end{equation}
admits three possible solutions, associated to the three 3-partitions $(\square; \bullet; \bullet)$, $(\bullet; \square; \bullet)$, $(\bullet; \bullet; \square)$ of $k=1$, with eigenvalues
\begin{equation}
\begin{split}
\mathcal{E}_1^{(3;(\square; \bullet; \bullet))} &= \alpha_1 (q + t^{-1} - qt^{-1}) + \alpha_2 + \alpha_3 
+ p\,\alpha_1 \dfrac{(1-q)(1-t)(q-t)(q\alpha_1 - t\alpha_2)(q\alpha_1 - t\alpha_3)}{t^4(\alpha_1 - \alpha_2)(\alpha_1 - \alpha_3)} + o(p^2) \\
\mathcal{E}_1^{(3;(\bullet; \square; \bullet))} &= \alpha_1 + \alpha_2 (q + t^{-1} - qt^{-1}) + \alpha_3 
+ p\, \alpha_2 \dfrac{(1-q)(1-t)(q-t)(q\alpha_2 - t\alpha_1)(q\alpha_2 - t\alpha_3)}{t^4(\alpha_2 - \alpha_1)(\alpha_2 - \alpha_3)} + o(p^2) \\ 
\mathcal{E}_1^{(3;(\bullet; \bullet; \square))} &= \alpha_1 + \alpha_2 + \alpha_3 (q + t^{-1} - qt^{-1}) 
+ p\, \alpha_3 \dfrac{(1-q)(1-t)(q-t)(q\alpha_3 - t\alpha_1)(q\alpha_3 - t\alpha_2)}{t^4(\alpha_3 - \alpha_1)(\alpha_3 - \alpha_2)} + o(p^2) \,.
\end{split}
\end{equation}

\subsection{$\Delta\text{ILW}_4$ Spectrum} \label{Sec:sec3.4}

In the $N=4$ case the first Hamiltonian reads
\begin{equation} 
\begin{split}
\widehat{\mathcal{H}}^{(4)}_1 = \Big[ & \alpha_1 \eta^{(1)}\left(z; p^{(1)} \right)
+ \alpha_2 \varphi_-^{(1)}\left((q^{-1}t)^{\frac{1}{4}}z; p^{(1)}\right) 
\eta^{(2)}\left((q^{-1}t)^{\frac{1}{2}}z; p^{(2)}\right) \\
& + \alpha_3 \varphi_-^{(1)}\left((q^{-1}t)^{\frac{1}{4}}z; p^{(1)}\right)
\varphi_-^{(2)}\left((q^{-1}t)^{\frac{3}{4}}z; p^{(2)}\right)
\eta^{(3)}\left((q^{-1}t) z; p^{(3)}\right) \\
& + \alpha_4 \varphi_-^{(1)}\left((q^{-1}t)^{\frac{1}{4}}z; p^{(1)}\right)
\varphi_-^{(2)}\left((q^{-1}t)^{\frac{3}{4}}z; p^{(2)}\right) \\
& \;\;\;\; \varphi_-^{(3)}\left((q^{-1}t)^{\frac{5}{4}}z; p^{(3)}\right)
\eta^{(4)}\left((q^{-1}t)^{\frac{3}{2}} z; p^{(4)}\right)
\Big]_1
\end{split}
\end{equation}
with $p^{(1)}=p(qt^{-1})^3$, $p^{(2)} = p(qt^{-1})^2$, $p^{(3)} = p(qt^{-1})$ and $p^{(4)} = p$.

\subsubsection{Zero solitons} \label{oscN4k0}

The only possible state with $k=0$ is the vacuum 
\begin{equation}
c_1 \vert 0 \rangle
\end{equation}
which corresponds to the only 4-partition $(\bullet; \bullet; \bullet; \bullet)$ of $k=0$.
The $\widehat{\mathcal{H}}_1^{(4)}$ action
\begin{equation}
\widehat{\mathcal{H}}_1^{(4)} c_1 \vert 0 \rangle = \left[\alpha_1 + \alpha_2 + \alpha_3 + \alpha_4 \right] c_1 \vert 0 \rangle = \mathcal{E}_1^{(4)} c_1 \vert 0 \rangle
\end{equation}
gives the vacuum energy
\begin{equation}
\mathcal{E}_1^{(4;(\bullet; \bullet; \bullet; \bullet))} = \alpha_1 + \alpha_2 + \alpha_3 + \alpha_4 \,.
\end{equation}

\subsubsection{One soliton} \label{oscN4k1}

A generic state with $k=1$ can be written as
\begin{equation}
(c_1 \lambda_{-1}^{(1)} + c_2 \lambda_{-1}^{(2)} + c_3 \lambda_{-1}^{(3)} + c_4 \lambda_{-1}^{(4)}) \vert 0 \rangle
\end{equation}
The eigenstate equation
\begin{equation}
\widehat{\mathcal{H}}_1^{(4)} (c_1 \lambda_{-1}^{(1)} + c_2 \lambda_{-1}^{(2)} + c_3 \lambda_{-1}^{(3)} + c_4 \lambda_{-1}^{(4)}) \vert 0 \rangle
= \mathcal{E}_1^{(4)} (c_1 \lambda_{-1}^{(1)} + c_2 \lambda_{-1}^{(2)} + c_3 \lambda_{-1}^{(3)} + c_4 \lambda_{-1}^{(4)}) \vert 0 \rangle
\end{equation}
admits three possible solutions, associated to the four 4-partitions $(\square; \bullet; \bullet; \bullet)$, $(\bullet; \square; \bullet; \bullet)$, $(\bullet; \bullet; \square; \bullet)$, $(\bullet; \bullet; \bullet; \square)$ of $k=1$, with eigenvalues
\begin{equation}
\begin{split}
\mathcal{E}_1^{(4;(\square; \bullet; \bullet; \bullet))} &= \alpha_1 (q + t^{-1} - qt^{-1}) + \alpha_2 + \alpha_3 + \alpha_4 \\
& + p\,\alpha_1 \dfrac{(1-q)(1-t)(q-t)(q\alpha_1 - t\alpha_2)(q\alpha_1 - t\alpha_3)(q\alpha_1 - t\alpha_4)}{t^5(\alpha_1 - \alpha_2)(\alpha_1 - \alpha_3)(\alpha_1 - \alpha_4)} + o(p^2) \\
\mathcal{E}_1^{(4;(\bullet; \square; \bullet; \bullet))} &= \alpha_1 + \alpha_2 (q + t^{-1} - qt^{-1}) + \alpha_3 + \alpha_4 \\
& + p\, \alpha_2 \dfrac{(1-q)(1-t)(q-t)(q\alpha_2 - t\alpha_1)(q\alpha_2 - t\alpha_3)(q\alpha_2 - t\alpha_4)}{t^5(\alpha_2 - \alpha_1)(\alpha_2 - \alpha_3)(\alpha_2 - \alpha_4)} + o(p^2) \\ 
\mathcal{E}_1^{(4;(\bullet; \bullet; \square; \bullet))} &= \alpha_1 + \alpha_2 + \alpha_3 (q + t^{-1} - qt^{-1}) + \alpha_4 \\
& + p\, \alpha_3 \dfrac{(1-q)(1-t)(q-t)(q\alpha_3 - t\alpha_1)(q\alpha_3 - t\alpha_2)(q\alpha_3 - t\alpha_4)}{t^5(\alpha_3 - \alpha_1)(\alpha_3 - \alpha_2)(\alpha_3 - \alpha_4)} + o(p^2) \\ 
\mathcal{E}_1^{(4;(\bullet; \bullet; \bullet; \square))} &= \alpha_1 + \alpha_2 + \alpha_3 + \alpha_4 (q + t^{-1} - qt^{-1}) \\
& + p\, \alpha_4 \dfrac{(1-q)(1-t)(q-t)(q\alpha_4 - t\alpha_1)(q\alpha_4 - t\alpha_2)(q\alpha_4 - t\alpha_3)}{t^5(\alpha_4 - \alpha_1)(\alpha_4 - \alpha_2)(\alpha_4 - \alpha_3)} + o(p^2) \,.
\end{split}
\end{equation}

\section{Bethe Ansatz equations for $\Delta\text{ILW}_N$ from 3d ADHM Theory} \label{Sec:sec4}
In the previous Section we explained how to compute the spectrum $\mathcal{E}^{(N)}_1$ of the first Hamiltonian \eqref{ham} of the quantum $\Delta\text{ILW}_N$ system by solving the associated eigenstate equation; in this Section we will show that the same spectrum can be obtained from the computation of a local observable in the 3d non-Abelian ADHM theory, generalizing what suggested in \cite{Koroteev:2015dja} for the Abelian case. 
%This follows naturally from the correspondence between the differential ($\text{ILW}_N$) version of our $\Delta\text{ILW}_N$ system and the 2d non-Abelian ADHM theory proposed in \cite{Ntalk,Otalk,NOinp,2013JHEP...11..155L,2014JHEP...07..141B,2015JHEP...02..150A,2015arXiv150507116B}. \\

As reviewed in Appendix \ref{appA}, the ADHM quiver theory plays a key role in the study of instantons in gauge theories -- its Higgs branch is isomorphic to the moduli space $\mathcal{M}_{k,N}$ of $k$ instantons in the $U(N)$ super Yang-Mills theory. The statement holds in any number of dimensions \cite{Tong:2005un}.  A 3d $\CN=2$ theory on $\mathbb{C} \times S^1_{\gamma}$ also admits a Coulomb branch which can be related to a quantum integrable trigonometric spin chain via Bethe/Gauge correspondence \cite{Nekrasov:2009uh,Nekrasov:2009ui,Gaiotto:2013bwa}. In particular, the equations determining the Coulomb branch supersymmetric vacua coincide with the Bethe Ansatz Equations of the associated spin chain (so there is a one-to-one correspondence between the supersymmetric vacua and the eigenstates of quantum Hamiltonians). Here we propose and give numerical evidence for the fact that the $\Delta\text{ILW}_N$ system is the integrable system associated to the 3d ADHM theory on $\mathbb{C} \times S^1_{\gamma}$. 

We proceed as follows. The Bethe Ansatz Equations determined by our ADHM theory read (see Appendix \ref{appA} for more details and the definition of the $q,t,\widetilde{p},\alpha_l,\sigma_s$ variables in terms of ADHM parameters)
\begin{equation}
\begin{split}
& \prod_{l=1}^N (\sigma_s \alpha_l^{-1} - 1) \prod_{\substack{t = 1 \\ t\neq s}}^k \dfrac{(\sigma_{s} - q \sigma_t) (\sigma_{s} - t^{-1} \sigma_t)}{( \sigma_{s} - \sigma_t) (\sigma_{s} - q t^{-1} \sigma_{t})} = \\
& = \widetilde{p} \, (-\sqrt{q t^{-1}})^N \, \prod_{l=1}^N (\sigma_s \alpha_l^{-1} - q^{-1} t) \prod_{\substack{t = 1 \\ t\neq s}}^k \dfrac{(\sigma_{s} - q^{-1} \sigma_t) (\sigma_{s} - t \sigma_t)}{(\sigma_{s} - \sigma_t) (\sigma_{s} - q^{-1} t \sigma_{t})}\,. \label{BAE}
\end{split}
\end{equation}
Here $\widetilde{p} = e^{-2\pi\xi}$ with $\xi$ Fayet-Iliopoulos parameter of our gauge theory.\footnote{In terms of Higgs branch target space, the parameter $\xi$ coincides with the K\"{a}hler modulus of $\mathcal{M}_{k,N}$. Moreover, $\widetilde{p}$ enters as the quantum deformation parameter in the equivariant quantum cohomology of the ADHM moduli space, or Hilbert scheme of $k$ points in the Abelian ($N=1$) case \cite{2004math.....11210O,2006math.....10129B,2011arXiv1106.3724C,Ciocan-Fontanine09quantumcohomology,2012arXiv1211.1287M,2014JHEP...01..038B,2015arXiv150507116B}.}
As we will see, $\widetilde{p}$ is related to the elliptic deformation parameter $p$ of $\text{eRS}_{N}$ and  $\Delta\text{ILW}_{N}$ via
\begin{equation}
\tilde{p} = p \,(-\sqrt{qt^{-1}})^{N}\,. \label{identification}
\end{equation}
One can show that at $\widetilde{p} = 0$ the solutions to equation \eqref{BAE} are labelled by $N$-partitions $\vec{\lambda} = (\lambda^{(1)}; \ldots; \lambda^{(N)})$ of $k$, and that this structure remains when the solutions are expanded in $\widetilde{p}$ small; this agrees with the fact that also the $\Delta\text{ILW}_N$ eigenstates can be put in one-to-one correspondence with $N$-partitions of $k$. \\
By extending the proposal in \cite{Koroteev:2015dja}, we suggest the vacuum expectation value of the equivariant Chern character of the universal $U(N)$ bundle over the instanton moduli space evaluated at $\vec{\lambda}$ to be the gauge theory  
observable corresponding to the eigenvalue of the first quantum $\Delta \text{ILW}_{N}$ Hamiltonian \eqref{ham}, i.e.
\begin{equation}
\begin{split}
\mathcal{E}^{(N; \vec{\lambda})}_1 & = \sum_{l=1}^N \left[ \alpha_l - (1-q)(1-t^{-1}) \sum_s \sigma_s^{(l)} \Big\vert_{\lambda^{(l)}} \right] \\
& = \sum_{l=1}^N \alpha_l - (1-q)(1-t^{-1}) \sum_s \sigma_s \Big\vert_{\vec{\lambda}}  \label{chernN}
\end{split}
\end{equation}
In the following we will check the validity of this proposal by revising the solitonic configurations from Section \ref{Sec:sec3}.

\subsection{$\Delta\text{ILW}_2$ Spectrum from 3d ADHM Theory} \label{Sec:sec4.1}

\subsubsection{Zero solitons} \label{ADHMN2k0}

When $k=0$ there are no equations; we can think of the associated ``solution'' as being the empty 2-partition $(\bullet; \bullet)$ of $k=0$. Formula \eqref{chernN} gives the energy
\begin{equation}
\mathcal{E}^{(2; (\bullet; \bullet))}_1 = \alpha_1 + \alpha_2
\end{equation}
which coincides with the result of Section \ref{oscN2k0}.

\subsubsection{One soliton} \label{ADHMN2k1}

When $k=1$ equations \eqref{BAE} admit two solutions, corresponding to the two 2-partitions $(\square; \bullet)$ and $(\bullet; \square)$ of $k=1$. Formula \eqref{chernN} gives the energies
\begin{equation}
\begin{split}
\mathcal{E}^{(2; (\square; \bullet))}_1 & \;=\; \alpha_1(q + t^{-1} - qt^{-1}) + \alpha_2 
+ \tilde{p} \, \alpha_1 \dfrac{(1-q)(1-t)(q-t)(q \alpha_1 - t \alpha_2)}{qt^2(\alpha_1 - \alpha_2)} \\
& + \tilde{p}^2 \alpha_1 \dfrac{(1-q)(1-t)(q-t)(q \alpha_1 - t \alpha_2)(q^2 \alpha_1^2 - (2q^2-qt+t^2)\alpha_1 \alpha_2 + qt\alpha_2^2)}{q^2t^3(\alpha_1 - \alpha_2)^3} + o(\tilde{p}^3) \\ 
\mathcal{E}^{(2; (\bullet; \square))}_1 & \;=\; \alpha_1 + \alpha_2(q + t^{-1} - qt^{-1}) 
+ \tilde{p} \, \alpha_2 \dfrac{(1-q)(1-t)(q-t)(q \alpha_2 - t \alpha_1)}{qt^2(\alpha_2 - \alpha_1)} \\
& + \tilde{p}^2 \alpha_2 \dfrac{(1-q)(1-t)(q-t)(q \alpha_2 - t \alpha_1)(q^2 \alpha_2^2 - (2q^2-qt+t^2)\alpha_1 \alpha_2 + qt\alpha_1^2)}{q^2t^3(\alpha_2 - \alpha_1)^3} + o(\tilde{p}^3) \,;
\end{split}
\end{equation}
these coincide with the results of Section \ref{oscN2k1} after identifying $\tilde{p} = p q t^{-1}$, as anticipated in \eqref{identification}.

\subsubsection{Two solitons} \label{ADHMN2k2}

When $k=2$ equations \eqref{BAE} admit five solutions, corresponding to the five 2-partitions $(\tiny{\yng(1,1)}; \bullet)$, $(\tiny{\yng(2)}; \bullet)$, $(\tiny{\yng(1,1)}; \bullet)$, $(\bullet; \tiny{\yng(2)})$, $(\tiny{\yng(1)}; \tiny{\yng(1)})$ of $k=2$. Formula \eqref{chernN} gives the energies
\begin{equation}
\begin{split}
\mathcal{E}^{(2; (\tiny{\yng(1,1)}; \bullet))}_1 & \;=\; \alpha_1(q^2 + t^{-1} - q^2t^{-1}) + \alpha_2 
+ \tilde{p} \, \alpha_1 \dfrac{(1-q^2)(1-t)^2(q-t)(q^2 \alpha_1 - t \alpha_2)}{t^2(1-qt)(q\alpha_1 - \alpha_2)} + o(\tilde{p}^2) \\ 
\mathcal{E}^{(2; (\tiny{\yng(2)}; \bullet))}_1 & \;=\; \alpha_1(q + t^{-2} - qt^{-2}) + \alpha_2 
+ \tilde{p} \, \alpha_1 \dfrac{(1-q)^2(1-t^2)(q-t)(q \alpha_1 - t^2 \alpha_2)}{qt^3(1-qt)(\alpha_1 - t\alpha_2)} + o(\tilde{p}^2) \\ 
\mathcal{E}^{(2; (\bullet; \tiny{\yng(1,1)}))}_1 & \;=\; \alpha_1 + \alpha_2(q^2 + t^{-1} - q^2t^{-1}) 
+ \tilde{p} \, \alpha_2 \dfrac{(1-q^2)(1-t)^2(q-t)(q^2 \alpha_2 - t \alpha_1)}{t^2(1-qt)(q\alpha_2 - \alpha_1)} + o(\tilde{p}^2) \\ 
\mathcal{E}^{(2; (\bullet; \tiny{\yng(2)}))}_1 & \;=\; \alpha_1 + \alpha_2(q + t^{-2} - qt^{-2}) 
+ \tilde{p} \, \alpha_2 \dfrac{(1-q)^2(1-t^2)(q-t)(q \alpha_2 - t^2 \alpha_1)}{qt^3(1-qt)(\alpha_2 - t\alpha_1)} + o(\tilde{p}^2) \\ 
\mathcal{E}^{(2; (\tiny{\yng(1)}; \tiny{\yng(1)}))}_1 & \;=\; (\alpha_1 + \alpha_2) (q + t^{-1} - qt^{-1}) \\
& + \tilde{p} \alpha_1 \dfrac{(1-q)(1-t)(q-t)(q\alpha_1 - \alpha_2)(t\alpha_1 - q\alpha_2)(\alpha_1 - t\alpha_2)}{qt^2(\alpha_1 - \alpha_2)(t\alpha_1 - \alpha_2)(\alpha_1 - q\alpha_2)} \\
& + \tilde{p} \alpha_2 \dfrac{(1-q)(1-t)(q-t)(q\alpha_2 - \alpha_1)(t\alpha_2 - q\alpha_1)(\alpha_2 - t\alpha_1)}{qt^2(\alpha_2 - \alpha_1)(t\alpha_2 - \alpha_1)(\alpha_2 - q\alpha_1)} + o(\tilde{p}^2) 
\end{split}
\end{equation}
which coincide with the results of section \ref{oscN2k2} after identifying $\tilde{p} = p q t^{-1}$.

\subsection{$\Delta\text{ILW}_3$ Spectrum from 3d ADHM Theory} \label{Sec:sec4.2}

\subsubsection{Zero solitons} \label{ADHMN3k0}

As before, for $k=0$ there are no equations and the associated ``solution'' corresponds to the empty 3-partition $(\bullet; \bullet; \bullet)$ of $k=0$. Formula \eqref{chernN} gives the energy
\begin{equation}
\mathcal{E}^{(3; (\bullet; \bullet; \bullet))}_1 = \alpha_1 + \alpha_2 + \alpha_3
\end{equation}
which coincides with the result of section \ref{oscN3k0}.

\subsubsection{One soliton} \label{ADHMN3k1}

When $k=1$ equations \eqref{BAE} admit three solutions, corresponding to the three 3-partitions $(\square; \bullet; \bullet)$, $(\bullet; \square; \bullet)$ and $(\bullet; \bullet; \square)$ of $k=1$. Formula \eqref{chernN} gives the energies
\begin{equation}
\begin{split}
\mathcal{E}^{(3; (\square; \bullet; \bullet))}_1 & \;=\; \alpha_1 (q + t^{-1} - qt^{-1}) + \alpha_2 + \alpha_3 \\
& - \dfrac{\tilde{p}}{\sqrt{qt^{-1}}}\, \alpha_1 \dfrac{(1-q)(1-t)(q-t)(q\alpha_1 - t\alpha_2)(q\alpha_1 - t\alpha_3)}{qt^3(\alpha_1 - \alpha_2)(\alpha_1 - \alpha_3)} + o(\tilde{p}^2) \\ 
\mathcal{E}^{(3; (\bullet; \square; \bullet))}_1 & \;=\; \alpha_1 + \alpha_2 (q + t^{-1} - qt^{-1}) + \alpha_3 \\
& - \dfrac{\tilde{p}}{\sqrt{qt^{-1}}}\, \alpha_2 \dfrac{(1-q)(1-t)(q-t)(q\alpha_2 - t\alpha_1)(q\alpha_2 - t\alpha_3)}{qt^3(\alpha_2 - \alpha_1)(\alpha_2 - \alpha_3)} + o(\tilde{p}^2) \\ 
\mathcal{E}^{(3; (\bullet; \bullet; \square))}_1 & \;=\; \alpha_1 + \alpha_2 + \alpha_3 (q + t^{-1} - qt^{-1}) \\
& - \dfrac{\tilde{p}}{\sqrt{qt^{-1}}}\, \alpha_3 \dfrac{(1-q)(1-t)(q-t)(q\alpha_3 - t\alpha_1)(q\alpha_3 - t\alpha_2)}{qt^3(\alpha_3 - \alpha_1)(\alpha_3 - \alpha_2)} + o(\tilde{p}^2) \,;
\end{split}
\end{equation}
these coincide with the results of section \ref{oscN3k1} after identifying $\tilde{p} = - p (q t^{-1})^{\frac{3}{2}}$.

\subsection{$\Delta\text{ILW}_4$ Spectrum from 3d ADHM Theory} \label{Sec:sec4.3}

\subsubsection{Zero solitons} \label{ADHMN4k0}

Once again, equations \eqref{BAE} reduce to nothing for $k=0$, and  the only ``solution'' corresponds to the empty 4-partition $(\bullet; \bullet; \bullet; \bullet)$ of $k=0$. Formula \eqref{chernN} gives the energy
\begin{equation}
\mathcal{E}^{(4; (\bullet; \bullet; \bullet))}_1 = \alpha_1 + \alpha_2 + \alpha_3 + \alpha_4
\end{equation}
which coincides with the result of section \ref{oscN4k0}.

\subsubsection{One soliton} \label{ADHMN4k1}

In the $k=1$ case equations \eqref{BAE} admit four solutions, corresponding to the four 4-partitions $(\square; \bullet; \bullet; \bullet)$, $(\bullet; \square; \bullet; \bullet)$, $(\bullet; \bullet; \square; \bullet)$ and $(\bullet; \bullet; \bullet; \square)$ of $k=1$. Formula \eqref{chernN} gives the energies
\begin{equation}
\begin{split}
\mathcal{E}^{(4; (\square; \bullet; \bullet; \bullet))}_1 & \;=\; \alpha_1 (q + t^{-1} - qt^{-1}) + \alpha_2 + \alpha_3 + \alpha_4 \\
& + \dfrac{\tilde{p}}{qt^{-1}}\, \alpha_1 \dfrac{(1-q)(1-t)(q-t)(q\alpha_1 - t\alpha_2)(q\alpha_1 - t\alpha_3)(q\alpha_1 - t\alpha_4)}{qt^4(\alpha_1 - \alpha_2)(\alpha_1 - \alpha_3)(\alpha_1 - \alpha_4)} + o(\tilde{p}^2) \\ 
\mathcal{E}^{(4; (\bullet; \square; \bullet; \bullet))}_1 & \;=\; \alpha_1 + \alpha_2 (q + t^{-1} - qt^{-1}) + \alpha_3 + \alpha_4 \\
& + \dfrac{\tilde{p}}{qt^{-1}}\, \alpha_2 \dfrac{(1-q)(1-t)(q-t)(q\alpha_2 - t\alpha_1)(q\alpha_2 - t\alpha_3)(q\alpha_2 - t\alpha_4)}{qt^4(\alpha_2 - \alpha_1)(\alpha_2 - \alpha_3)(\alpha_2 - \alpha_4)} + o(\tilde{p}^2) \\ 
\mathcal{E}^{(4; (\bullet; \bullet; \square; \bullet))}_1 & \;=\; \alpha_1 + \alpha_2 + \alpha_3 (q + t^{-1} - qt^{-1}) + \alpha_4 \\
& + \dfrac{\tilde{p}}{qt^{-1}}\, \alpha_3 \dfrac{(1-q)(1-t)(q-t)(q\alpha_3 - t\alpha_1)(q\alpha_3 - t\alpha_2)(q\alpha_3 - t\alpha_4)}{qt^4(\alpha_3 - \alpha_1)(\alpha_3 - \alpha_2)(\alpha_3 - \alpha_4)} + o(\tilde{p}^2) \\
\mathcal{E}^{(4; (\bullet; \bullet; \bullet; \square))}_1 & \;=\; \alpha_1 + \alpha_2 + \alpha_3 + \alpha_4 (q + t^{-1} - qt^{-1}) \\
& + \dfrac{\tilde{p}}{qt^{-1}}\, \alpha_4 \dfrac{(1-q)(1-t)(q-t)(q\alpha_4 - t\alpha_1)(q\alpha_4 - t\alpha_2)(q\alpha_4 - t\alpha_3)}{qt^4(\alpha_4 - \alpha_1)(\alpha_4 - \alpha_2)(\alpha_4 - \alpha_3)} + o(\tilde{p}^2) \,;
\end{split}
\end{equation}
these coincide with the results of section \ref{oscN4k1} after identifying $\tilde{p} = p (q t^{-1})^{2}$.

\section{$\Delta \text{ILW}_{N}$ as Large-$n$ Limit of $N$ Coupled Ruijsenaars-Schneider Models} \label{Sec:sec5}
Having discussed in the previous sections the computation of the $\Delta\text{ILW}_N$ spectrum, we would now like to understand how to obtain the spectrum for the $\text{eRS}_N$ system and show that this reduces to the $\Delta\text{ILW}_N$ one in the $n \rightarrow \infty$ limit according to \eqref{elllimN}. 

As we recalled in Section \ref{Sec:sec2.1}, when $N=1$ the eigenvalue $E_{eRS}^{(\lambda;n)}(p)$ of the first $n$-particles eRS Hamiltonian relative to an eigenfunction labelled by a partition $\lambda$ of $k$ (the degree of the Macdonald polynomial) coincides with the vacuum expectation value of the Wilson loop in the fundamental representation of the 5d $\mathcal{N}=1^*$ $U(n)$ theory evaluated at the supersymmetric vacuum associated to $\lambda$. In formulas
\begin{equation}
E_{eRS}^{(\lambda;n)}(p) \;=\; \left\langle W_{\square}^{SU(n)} \right\rangle\Big\vert_{\lambda} \;=\; \left\langle W_{\square}^{U(n)} \right\rangle \Big/ \left\langle W_{\square}^{U(1)} \right\rangle\Big\vert_{\lambda} \label{WSU5}
\end{equation}
with
\begin{equation}
\left\langle W_{\square}^{U(n)} \right\rangle = \sum_{a=1}^n \mu_a - Q \dfrac{(q-t)(1-t)}{q t^n} \sum_{a=1}^n \mu_a
\prod_{\substack{b=1 \\ b \neq a}}^n \dfrac{(\mu_a - t \mu_b)(t \mu_a - q \mu_b)}{(\mu_a - \mu_b)(\mu_a - q \mu_b)} + o(Q^2)\,, \label{WU5}
\end{equation}
\begin{equation}
\left\langle W_{\square}^{U(1)} \right\rangle = \dfrac{(Qt^{-1};Q)_{\infty}(Qtq^{-1};Q)_{\infty}}{(Q;Q)_{\infty}(Qq^{-1};Q)_{\infty}}\,.
\end{equation}
The exponentiated 5d coupling $Q = e^{-8\pi^2 \gamma/g^2_{YM}}$ is proportional to elliptic parameter deformation $p$. Evaluation of \eqref{WSU5} at the partition $\lambda = (\lambda_1, \ldots, \lambda_n)$ of $k$ means fixing the $\mu_a$ parameters according to
\begin{equation}
\mu_a = q^{\lambda_a}t^{n-a} \;\;\;,\;\;\; a = 1, \ldots, n \,. \label{locus5}
\end{equation}
Here we propose \eqref{WSU5} to also be valid in the $N>1$ case, if we replace $U(n)$ by $U(Nn)$. In more details, we consider the case in which the $Nn$ particles of the $N$ coupled eRS systems are split into $N$ sets of $n$ particles each. Then the eigenfunctions (generalized Macdonald polynomials of degree $k$ in the trigonometric limit) will be in the one-to-one correspondence with $N$-partitions $\vec{\lambda} = (\lambda^{(1)}; \ldots; \lambda^{(N)})$ of $k$ where each partition $\lambda^{(l)}$ is of length $n$ (which is just the number of particles in each of the $N$ eRS systems); at the level of supersymmetric vacuum, this corresponds to splitting the 5d Coulomb branch parameters $\mu_a$, $a = 1, \ldots, Nn$ into $N$ sets $\mu_a^{(l)}$, $l = 1, \ldots, N$, $a = 1, \ldots, n$ and fix them to
\begin{equation}
\mu_a^{(l)} = \widetilde{\alpha}_l q^{\lambda_a}t^{n-a} \;\;\;,\;\;\; l = 1, \ldots, N \;\;\;,\;\;\; a = 1, \ldots, n \,. \label{locus5N}
\end{equation}
Alternatively we can think of a large Young tableaux $\Lambda$, whose column numbers run $1\dots nN$, which can be built as shown the figure below.
\begin{figure}
\label{fig:bigtableux}
\begin{center}
\includegraphics[scale=0.35]{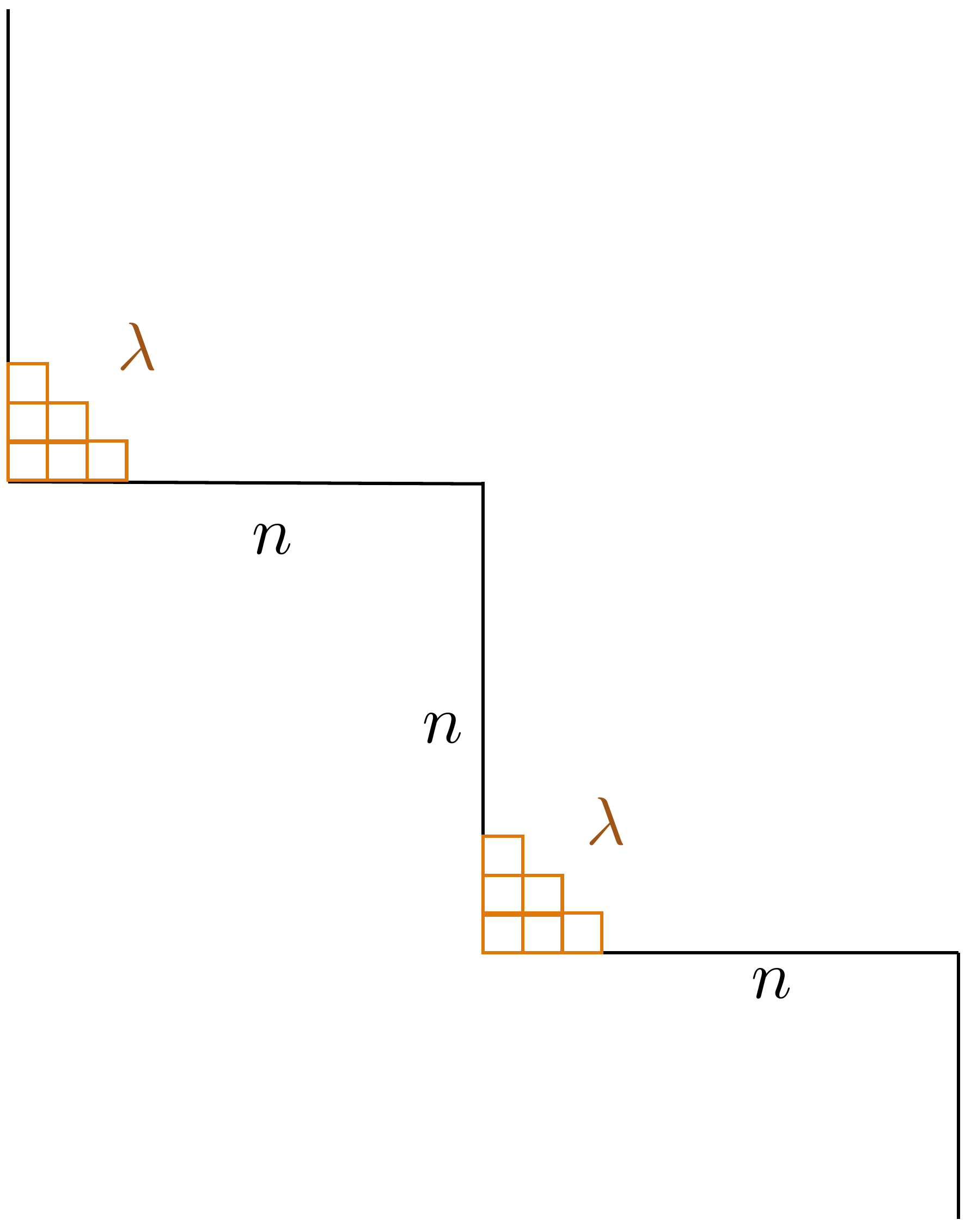}
\caption{Large tableaux $\Lambda$ which specifies Higgs branch condition (\ref{eq:HiggsILWNL}). It represents a ladder with step length and height equal to $n$ with tableaux $\lambda$ inserted in corners. }
\end{center} 
\end{figure} 
Then the Higgsing condition reads
\begin{equation}
\label{eq:HiggsILWNL}
\mu_I = \widetilde{\alpha}_I q^{\Lambda_I} t^{nN-I}\,.
\end{equation}
Our proposal 
\begin{equation}
E_{eRS_N}^{(\vec{\lambda};Nn)}(p) \;=\; \left\langle W_{\square}^{SU(Nn)} \right\rangle\Big\vert_{\vec{\lambda}} \;=\; \left\langle W_{\square}^{U(Nn)} \right\rangle \Big/ \left\langle W_{\square}^{U(1)} \right\rangle\Big\vert_{\vec{\lambda}} \label{WSU5N}
\end{equation}
can be immediately verified in the trigonometric case: in fact when $p=0$ we have
\begin{equation}
E_{tRS_N}^{(\vec{\lambda};Nn)} \;=\; \sum_{l=1}^N \sum_{a=1}^n \widetilde{\alpha}_l q^{\lambda_a^{(l)}}t^{n-a} = 
t^{n-1} \sum_{l=1}^N \sum_{a=1}^n \widetilde{\alpha}_l (q^{\lambda_a^{(l)}} - 1) t^{1-a} + t^{n-1} \dfrac{1-t^{-n}}{1-t^{-1}} \sum_{l=1}^N \widetilde{\alpha}_l 
\end{equation}
therefore (see equation \eqref{questa})
\begin{equation}
t^{-n}\sum_{l=1}^N \widetilde{\alpha}_l + t^{-n+1}(1-t^{-1}) E_{tRS_N}^{(\vec{\lambda};Nn)} = 
\sum_{l=1}^N \widetilde{\alpha}_l + (1-t^{-1}) \sum_{l=1}^N \sum_{a=1}^n \widetilde{\alpha}_l (q^{\lambda_a^{(l)}} - 1) t^{1-a}
\end{equation}
which exactly coincides with the $\Delta\text{BO}_N$ spectrum \eqref{appDBON} for $\widetilde{\alpha}_l = \alpha_l$. Note that we keep $\widetilde{\alpha}_l$ distinct from $\alpha_l$, because in more general situations, these parameters will only be proportional to each other: we have in mind cases in which one splits the set of $nN$ particles in $N$ sets with different number of particles (or equivalently, when tableaux $\lambda$ inside the corners of $\Lambda$). Since we have defined the $\text{tRS}_{N}$ system starting from the $\Delta\text{BO}_N$ one, and \eqref{WSU5N} reproduces the $\text{tRS}_{N}$ spectrum implied by our definition, we conclude that our proposal \eqref{WSU5N} is correct. The analogy with the $N=1$ case gives us confidence on the validity of \eqref{WSU5N} even in the elliptic case; in fact later in this section we will check explicitly in many cases that the expected large $n$ relation \eqref{elllimN}, i.e.
\begin{equation}
\mathcal{E}^{(N; \vec{\lambda})}_1 = 1 - (1-q)(1-t^{-1})\text{Tr}\, \sigma \big\vert_{\vec{\lambda}} = \lim_{n \rightarrow \infty} \left[ t^{-n+1}(1-t^{-1}) \left\langle W_{\square}^{U(Nn)} \right\rangle \right] \Big\vert_{\vec{\lambda}} \,, \label{elllimN5}
\end{equation}
holds true, with the identification
\begin{equation}
Q = (- \sqrt{qt^{-1}})^N \widetilde{p} = (qt^{-1})^N p \,.
\end{equation}

At this moment we can generalize the statements which we have made in \secref{Sec:sec2.4}. Thus we show that the large-$n$ limit of the $eRS_N$ model leads us to the spin chain which is described by the twisted chiral ring relations of ADHM quiver gauge theory. The latter, according to the gauge/gravity duality, coincide with the Bethe Ansatz equations for $\widehat{A}_0$ XXZ spinchain on $N$ sites with impurities $\alpha_1,\dots,\alpha_N$ with the number of excitations provided by the quantization condition \eqref{locus5N}.

We will now proceed to check \eqref{elllimN5} in the several topological sectors as follows: at fixed $N$ we first fix $k$, then we compute $t^{-n+1}(1-t^{-1}) \left\langle W_{\square}^{U(Nn)} \right \rangle$ for all possible partitions of $k$ keeping $n$ generic, and finally we take the limit $n \rightarrow \infty$ and check with the already computed $\Delta\text{ILW}_N$ results. In order to lighten the text we will use a simplified notation for our partitions of length $n$, in which all non-written entries are zero. For example 
\begin{equation}
\begin{split}
& (\bullet) \;\;\; \text{corresponds to} \;\;\; (\underbrace{0, \ldots, 0}_n) \\
& (\square) \;\;\; \text{corresponds to} \;\;\; (1,\underbrace{0,\ldots, 0}_{n-1}) \\
& ({\tiny{\yng(1,1)}}) \;\;\; \text{corresponds to} \;\;\; (2,\underbrace{0,\ldots, 0}_{n-1}) \\
& ({\tiny{\yng(2)}}) \;\;\; \text{corresponds to} \;\;\; (1, 1, \underbrace{0,\ldots, 0}_{n-2}) \,.
\end{split}
\end{equation}
We have actually used the same convention in the previous Sections, where we were working with partitions of length $k$.

\subsection{$\Delta\text{ILW}_2$ Spectrum from $\text{eRS}_2$ Spectrum} \label{Sec:sec5.1}
\subsubsection{Zero solitons}
At $k=0$ we only have the empty 2-partition $(\bullet; \bullet)$ of $k$. From our formula we obtain
\begin{equation}
\begin{split}
& t^{-n+1}(1-t^{-1}) \left\langle W_{\square}^{U(2n)} \right \rangle \Big\vert_{(\bullet; \bullet)} = \left(\widetilde{\alpha}_1 + \widetilde{\alpha}_2 \right) \left(1-t^{-n}\right) \\
& + Q \widetilde{\alpha}_1 t^{-n} \dfrac{(1-q)(1-t)(q-t)\left(1-t^{-n}\right)}{q^3 \left(1- q^{-1} t^{1-n}\right)} 
\dfrac{(t \widetilde{\alpha}_1 - q \widetilde{\alpha}_2)(\widetilde{\alpha}_2 - t^{-n} \widetilde{\alpha}_1)}{(\widetilde{\alpha}_1 - \widetilde{\alpha}_2)(\widetilde{\alpha}_2 - q^{-1} t^{1-n} \widetilde{\alpha}_1)} \\
& + Q \widetilde{\alpha}_2 t^{-n} \dfrac{(1-q)(1-t)(q-t)\left(1-t^{-n}\right)}{q^3 \left(1- q^{-1} t^{1-n}\right)}  \dfrac{(t \widetilde{\alpha}_2 - q \widetilde{\alpha}_1)(\widetilde{\alpha}_1 - t^{-n} \widetilde{\alpha}_2)}{(\widetilde{\alpha}_2 - \widetilde{\alpha}_1)(\widetilde{\alpha}_1 - q^{-1} t^{1-n} \widetilde{\alpha}_2)} + o(Q^2)
\end{split}
\end{equation}
which, in the limit $n \rightarrow \infty$, reduces to
\begin{equation}
\widetilde{\alpha}_1 + \widetilde{\alpha}_2 + o(Q^2) \,.
\end{equation}
This immediately matches the results in Sections \ref{oscN2k0} and \ref{ADHMN2k0} for $\widetilde{\alpha}_{1,2} = \alpha_{1,2}$.

\subsubsection{One soliton}
When $k=1$ we have the two 2-partitions $(\square; \bullet)$, $(\bullet; \square)$ of $k$; our formula gives:
\begin{itemize}
\item Partition $(\square; \bullet)$:
\begin{equation}
\begin{split}
& t^{-n+1}(1-t^{-1}) \left\langle W_{\square}^{U(2n)} \right \rangle \Big\vert_{(\tiny{\yng(1)}; \bullet)} = \\
& \widetilde{\alpha}_1 (q-1)(1-t^{-1}) + \left(\widetilde{\alpha}_1 + \widetilde{\alpha}_2 \right) \left(1-t^{-n}\right) \\
& + Q \widetilde{\alpha}_1 \dfrac{(1-q)(q-t)(1-t^{-1})(1-t^{-n})}{q^3(1-q^{-1}t^{1-n})}
\dfrac{(\widetilde{\alpha}_1 - t^{-n}\widetilde{\alpha}_2)(q\widetilde{\alpha}_1 - t\widetilde{\alpha}_2)}{(\widetilde{\alpha}_2 - \widetilde{\alpha}_1)(\widetilde{\alpha}_1 - q^{-1}t^{1-n}\widetilde{\alpha}_2)} \\
& + Q \widetilde{\alpha}_1 t^{1-n} \dfrac{(1-q)(q-t)(1-t^{-1})(1-t^{1-n})(1-q^{-1}t^{-n})(1-q^{-2}t^{2-n})}{q^4(1-q^{-1}t^{1-n})(1-q^{-1}t^{2-n})(1-q^{-2}t^{1-n})} \\
& \times \dfrac{(\widetilde{\alpha}_2 - t^{-n}\widetilde{\alpha}_1)(q\widetilde{\alpha}_2 - t\widetilde{\alpha}_1)}{(\widetilde{\alpha}_1 - \widetilde{\alpha}_2)(\widetilde{\alpha}_2 - q^{-1}t^{1-n}\widetilde{\alpha}_1)} \\
& + o(Q^2) \label{N2k10} \,;
\end{split}
\end{equation}
in the limit $n \rightarrow \infty$, this reduces to
\begin{equation}
\widetilde{\alpha}_1 (q + t^{-1} - q t^{-1}) + \widetilde{\alpha}_2 + Q \widetilde{\alpha}_1 \dfrac{(1-q)(1-t)(q-t)(q \widetilde{\alpha}_1 - t \widetilde{\alpha}_2)}{q^2 t (\widetilde{\alpha}_1 - \widetilde{\alpha}_2)} + o(Q^2) 
\end{equation}
which matches the results in Sections \ref{oscN2k1} and \ref{ADHMN2k1} for $\widetilde{\alpha}_{1,2} = \alpha_{1,2}$ and $Q = p(qt^{-1})^2$.

\item Partition $(\bullet; \square)$:

Can be obtained from \eqref{N2k10} by permutation of the $\widetilde{\alpha}_a$ parameters. In the limit $n \rightarrow \infty$, this reduces to
\begin{equation}
\widetilde{\alpha}_1 + \widetilde{\alpha}_2 (q + t^{-1} - q t^{-1}) + Q \widetilde{\alpha}_2 \dfrac{(1-q)(1-t)(q-t)(q \widetilde{\alpha}_2 - t \widetilde{\alpha}_1)}{q^2 t (\widetilde{\alpha}_2 - \widetilde{\alpha}_1)} + o(Q^2)
\end{equation}
which matches the results in Sections \ref{oscN2k1} and \ref{ADHMN2k1} for $\widetilde{\alpha}_{1,2} = \alpha_{1,2}$ and $Q = p(qt^{-1})^2$.

\end{itemize}

\subsubsection{Two solitons}

When $k=2$ we have the five 2-partitions $(\tiny{\yng(1,1)}; \bullet)$, $(\tiny{\yng(2)}; \bullet)$, $(\bullet; \tiny{\yng(1,1)})$, $(\bullet; \tiny{\yng(2)})$, $(\tiny{\yng(1)}; \tiny{\yng(1)})$ of $k$; our formula gives:

\begin{itemize}
\item Partition $(\tiny{\yng(1,1)}; \bullet)$:
\begin{equation}
\begin{split}
& t^{-n+1}(1-t^{-1}) \left\langle W_{\square}^{U(2n)} \right \rangle \Big\vert_{(\tiny{\yng(1,1)}; \bullet)} = \\
& \widetilde{\alpha}_1 (q^2-1)(1-t^{-1}) + \left(\widetilde{\alpha}_1 + \widetilde{\alpha}_2 \right) \left(1-t^{-n}\right) \\
& + Q \widetilde{\alpha}_1 \dfrac{(1-q^2)(1-t^2)(q-t)(1-q^{-1}t^{-n})}{qt(1-qt)(1-q^{-2}t^{1-n})} 
\dfrac{(\widetilde{\alpha}_1 - q^{-1}t^{-n}\widetilde{\alpha}_2)(q^2 \widetilde{\alpha}_1 - t \widetilde{\alpha}_2)}{(q \widetilde{\alpha}_1 - \widetilde{\alpha}_2)(\widetilde{\alpha}_1 - q^{-2}t^{1-n} \widetilde{\alpha}_2)} \\
& + Q \widetilde{\alpha}_1 t^{1-n} \dfrac{(1-q)(1-t^{-1})(q-t)(1-t^{1-n})(1-q^{-2}t^{-n})(1-q^{-3}t^{2-n})}{q^3(1-q^{-1}t^{2-n})(1-q^{-2}t^{1-n})(1-q^{-3}t^{1-n})} \\
& \times \dfrac{(\widetilde{\alpha}_2 - t^{-n} \widetilde{\alpha}_1)(q \widetilde{\alpha}_2 - t \widetilde{\alpha}_1)}{(\widetilde{\alpha}_1 - \widetilde{\alpha}_2)(\widetilde{\alpha}_2 - q^{-1}t^{1-n}\widetilde{\alpha}_1)} \\
& + Q \widetilde{\alpha}_2 t^{1-n} \dfrac{(1-q)(1-t^{-1})(q-t)(1-t^{-n})}{q^3(1-q^{-1}t^{1-n})} \\
& \times \dfrac{(q\widetilde{\alpha}_1 - t\widetilde{\alpha}_2)(\widetilde{\alpha}_1 - t^{1-n}\widetilde{\alpha}_2)(\widetilde{\alpha}_1 - q^{-2}t^{-n}\widetilde{\alpha}_2)(\widetilde{\alpha}_1 - q^{-3}t^{2-n}\widetilde{\alpha}_2)}{(\widetilde{\alpha}_2 - \widetilde{\alpha}_1)(\widetilde{\alpha}_1 - q^{-1}t^{2-n}\widetilde{\alpha}_2)(\widetilde{\alpha}_1 - q^{-2}t^{1-n}\widetilde{\alpha}_2)(\widetilde{\alpha}_1 - q^{-3}t^{1-n}\widetilde{\alpha}_2)} \\
& + o(Q^2) \,;
\label{N2k20}
\end{split}
\end{equation}
in the limit $n \rightarrow \infty$, this reduces to
\begin{equation}
\widetilde{\alpha}_1 (q^2 + t^{-1} - q^2 t^{-1}) + \widetilde{\alpha}_2 + Q \widetilde{\alpha}_1 \dfrac{(1-q^2)(1-t)^2(q-t)(q^2 \widetilde{\alpha}_1 - t \widetilde{\alpha}_2)}{q t (1-qt)(q\widetilde{\alpha}_1 - \widetilde{\alpha}_2)} + o(Q^2) 
\end{equation}
which matches the results in Sections \ref{oscN2k2} and \ref{ADHMN2k2} for $\widetilde{\alpha}_{1,2} = \alpha_{1,2}$ and $Q = p(qt^{-1})^2$.

\item Partition $(\tiny{\yng(2)}; \bullet)$:
\begin{equation}
\begin{split}
& t^{-n+1}(1-t^{-1}) \left\langle W_{\square}^{U(2n)} \right \rangle \Big\vert_{(\tiny{\yng(2)}; \bullet)} = \\
& \widetilde{\alpha}_1 (q-1)(1-t^{-2}) + \left(\widetilde{\alpha}_1 + \widetilde{\alpha}_2 \right) \left(1-t^{-n}\right) \\
& - Q \widetilde{\alpha}_1 \dfrac{(1-q)^2(1-t^{-2})(q-t)(1-t^{1-n})}{q^2t(1-qt)(1-q^{-1}t^{2-n})} 
\dfrac{(\widetilde{\alpha}_1 - t^{1-n}\widetilde{\alpha}_2)(q \widetilde{\alpha}_1 - t^2 \widetilde{\alpha}_2)}{(t^{-1} \widetilde{\alpha}_1 - \widetilde{\alpha}_2)(\widetilde{\alpha}_1 - q^{-1}t^{2-n} \widetilde{\alpha}_2)} \\
& + Q \widetilde{\alpha}_1 t^{1-n} \dfrac{(1-q)(1-t^{-1})(q-t)(1-t^{2-n})(1-q^{-1}t^{-n})(1-q^{-2}t^{3-n})}{q^3(1-q^{-1}t^{2-n})(1-q^{-1}t^{3-n})(1-q^{-2}t^{1-n})} \\
& \times \dfrac{(\widetilde{\alpha}_2 - t^{-n} \widetilde{\alpha}_1)(q \widetilde{\alpha}_2 - t \widetilde{\alpha}_1)}{(\widetilde{\alpha}_1 - \widetilde{\alpha}_2)(\widetilde{\alpha}_2 - q^{-1}t^{1-n}\widetilde{\alpha}_1)} \\
& + Q \widetilde{\alpha}_2 t^{1-n} \dfrac{(1-q)(1-t^{-1})(q-t)(1-t^{-n})}{q^3(1-q^{-1}t^{1-n})} \\
& \times \dfrac{(q\widetilde{\alpha}_1 - t\widetilde{\alpha}_2)(\widetilde{\alpha}_1 - t^{2-n}\widetilde{\alpha}_2)(\widetilde{\alpha}_1 - q^{-1}t^{-n}\widetilde{\alpha}_2)(\widetilde{\alpha}_1 - q^{-2}t^{3-n}\widetilde{\alpha}_2)}{(\widetilde{\alpha}_2 - \widetilde{\alpha}_1)(\widetilde{\alpha}_1 - q^{-1}t^{2-n}\widetilde{\alpha}_2)(\widetilde{\alpha}_1 - q^{-1}t^{3-n}\widetilde{\alpha}_2)(\widetilde{\alpha}_1 - q^{-2}t^{1-n}\widetilde{\alpha}_2)} \\
& + o(Q^2) \,;
\label{N2k11}
\end{split}
\end{equation}
in the limit $n \rightarrow \infty$, this reduces to
\begin{equation}
\widetilde{\alpha}_1 (q + t^{-2} - q t^{-2}) + \widetilde{\alpha}_2 + Q \widetilde{\alpha}_1 \dfrac{(1-q)^2(1-t^2)(q-t)(q \widetilde{\alpha}_1 - t^2 \widetilde{\alpha}_2)}{q^2 t^2 (1-qt)(\widetilde{\alpha}_1 - t \widetilde{\alpha}_2)} + o(Q^2) 
\end{equation}
which matches the results in Sections \ref{oscN2k2} and \ref{ADHMN2k2} for $\widetilde{\alpha}_{1,2} = \alpha_{1,2}$ and $Q = p(qt^{-1})^2$.

\item Partition $(\bullet; \tiny{\yng(1,1)})$:

Can be obtained from \eqref{N2k20} by permutation of the $\widetilde{\alpha}_a$ parameters. In the limit $n \rightarrow \infty$, this reduces to
\begin{equation}
\widetilde{\alpha}_2 (q^2 + t^{-1} - q^2 t^{-1}) + \widetilde{\alpha}_1 + Q \widetilde{\alpha}_2 \dfrac{(1-q^2)(1-t)^2(q-t)(q^2 \widetilde{\alpha}_2 - t \widetilde{\alpha}_1)}{q t (1-qt)(q\widetilde{\alpha}_2 - \widetilde{\alpha}_1)} + o(Q^2) 
\end{equation}
which matches the results in Sections \ref{oscN2k2} and \ref{ADHMN2k2} for $\widetilde{\alpha}_{1,2} = \alpha_{1,2}$ and $Q = p(qt^{-1})^2$.

\item Partition $(\bullet; \tiny{\yng(2)})$:

Can be obtained from \eqref{N2k11} by permutation of the $\widetilde{\alpha}_a$ parameters. In the limit $n \rightarrow \infty$, this reduces to
\begin{equation}
\widetilde{\alpha}_2 (q + t^{-2} - q t^{-2}) + \widetilde{\alpha}_1 + Q \widetilde{\alpha}_2 \dfrac{(1-q)^2(1-t^2)(q-t)(q \widetilde{\alpha}_2 - t^2 \widetilde{\alpha}_1)}{q^2 t^2 (1-qt)(\widetilde{\alpha}_2 - t \widetilde{\alpha}_1)} + o(Q^2) 
\end{equation}
which matches the results in Sections \ref{oscN2k2} and \ref{ADHMN2k2} for $\widetilde{\alpha}_{1,2} = \alpha_{1,2}$ and $Q = p(qt^{-1})^2$.

\item Partition $(\tiny{\yng(1)}; \tiny{\yng(1)})$:
\begin{equation}
\begin{split}
& t^{-n+1}(1-t^{-1}) \left\langle W_{\square}^{U(2n)} \right \rangle \Big\vert_{(\tiny{\yng(1)}; \tiny{\yng(1)})} = \\
& \left(\widetilde{\alpha}_1 + \widetilde{\alpha}_2 \right) \left[(q-1)(1-t^{-1}) + \left(1-t^{-n}\right) \right] \\
& - Q \sum_{a=1}^2 \widetilde{\alpha}_a \dfrac{(1-q)(1-t^{-1})(q-t)(1-t^{-n})}{q^2(1-q^{-1}t^{1-n})} 
\prod_{\substack{b=1 \\ b\neq a}}^2
\dfrac{(\widetilde{\alpha}_a - t^{-n}\widetilde{\alpha}_b)(q \widetilde{\alpha}_a - \widetilde{\alpha}_b)(q \widetilde{\alpha}_b - t \widetilde{\alpha}_a)(t \widetilde{\alpha}_b - \widetilde{\alpha}_a)}{(\widetilde{\alpha}_a - q^{-1}t^{1-n} \widetilde{\alpha}_b)(\widetilde{\alpha}_b - \widetilde{\alpha}_a)(t\widetilde{\alpha}_a - \widetilde{\alpha}_b)(q\widetilde{\alpha}_b - \widetilde{\alpha}_a)} \\
& + Q \sum_{a=1}^2 \widetilde{\alpha}_a t^{1-n} \dfrac{(1-q)(1-t^{-1})(q-t)(1-t^{1-n})(1-q^{-1}t^{-n})(1-q^{-2}t^{2-n})}{q^3(1-q^{-1}t^{1-n})(1-q^{-1}t^{2-n})(1-q^{-2}t^{1-n})} \\
& \times \prod_{\substack{b=1 \\ b\neq a}}^2 \dfrac{(\widetilde{\alpha}_b - t^{1-n} \widetilde{\alpha}_a)(\widetilde{\alpha}_b - q^{-1}t^{-n} \widetilde{\alpha}_a)(\widetilde{\alpha}_b - q^{-2}t^{2-n} \widetilde{\alpha}_a)(q \widetilde{\alpha}_b - t \widetilde{\alpha}_a)}{(\widetilde{\alpha}_a - \widetilde{\alpha}_b)(\widetilde{\alpha}_b - q^{-1}t^{1-n}\widetilde{\alpha}_a)(\widetilde{\alpha}_b - q^{-1}t^{2-n}\widetilde{\alpha}_a)(\widetilde{\alpha}_b - q^{-2}t^{1-n}\widetilde{\alpha}_a)} \\
& + o(Q^2) \,;
\end{split}
\end{equation}
in the limit $n \rightarrow \infty$, this reduces to
\begin{equation}
\begin{split}
& (\widetilde{\alpha}_1 + \widetilde{\alpha}_2) (q + t^{-1} - q t^{-1}) + Q \widetilde{\alpha}_1 
\dfrac{(1-q)(1-t)(q-t)(q \widetilde{\alpha}_1 - \widetilde{\alpha}_2)(t \widetilde{\alpha}_1 - q \widetilde{\alpha}_2)(\widetilde{\alpha}_1 - t \widetilde{\alpha}_2)}{q^2 t(\widetilde{\alpha}_1 - \widetilde{\alpha}_2)(t\widetilde{\alpha}_1 - \widetilde{\alpha}_2)(\widetilde{\alpha}_1 - q\widetilde{\alpha}_2)} \\
& + Q \widetilde{\alpha}_2 \dfrac{(1-q)(1-t)(q-t)(q \widetilde{\alpha}_2 - \widetilde{\alpha}_1)(t \widetilde{\alpha}_2 - q \widetilde{\alpha}_1)(\widetilde{\alpha}_2 - t \widetilde{\alpha}_1)}{q^2 t(\widetilde{\alpha}_2 - \widetilde{\alpha}_1)(t\widetilde{\alpha}_2 - \widetilde{\alpha}_1)(\widetilde{\alpha}_2 - q\widetilde{\alpha}_1)} \\
& + o(Q^2) 
\end{split}
\end{equation}
which matches the results in Sections \ref{oscN2k2} and \ref{ADHMN2k2} for $\widetilde{\alpha}_{1,2} = \alpha_{1,2}$ and $Q = p(qt^{-1})^2$.

\end{itemize}

\subsection{$\Delta\text{ILW}_3$ Spectrum from $\text{eRS}_3$ Spectrum} \label{Sec:sec5.2}

\subsubsection{Zero solitons}

For $k=0$ we only have the empty 3-partition $(\bullet; \bullet; \bullet)$ of $k$. From our formula we obtain
\begin{equation}
\begin{split}
& t^{-n+1}(1-t^{-1}) \left\langle W_{\square}^{U(3n)} \right \rangle \Big\vert_{(\bullet; \bullet; \bullet)} = \left(\widetilde{\alpha}_1 + \widetilde{\alpha}_2 + \widetilde{\alpha}_3 \right) \left(1-t^{-n}\right) \\
& + Q t^{-n} \dfrac{(1-q)(1-t)(q-t)\left(1-t^{-n}\right)}{q^4 \left(1- q^{-1} t^{1-n}\right)} 
\sum_{a=1}^3 \widetilde{\alpha}_a \prod_{\substack{b=1 \\ b \neq a}}^3 \dfrac{(t \widetilde{\alpha}_a - q \widetilde{\alpha}_b)(\widetilde{\alpha}_b - t^{-n} \widetilde{\alpha}_a)}{(\widetilde{\alpha}_a - \widetilde{\alpha}_b)(\widetilde{\alpha}_b - q^{-1} t^{1-n} \widetilde{\alpha}_a)} \\
& + o(Q^2)
\end{split}
\end{equation}
which, in the limit $n \rightarrow \infty$, reduces to
\begin{equation}
\widetilde{\alpha}_1 + \widetilde{\alpha}_2 + \widetilde{\alpha}_3 + o(Q^2) \,.
\end{equation}
This immediately matches the results in Sections \ref{oscN3k0} and \ref{ADHMN3k0} for $\widetilde{\alpha}_{1,2,3} = \alpha_{1,2,3}$.

\subsubsection{One soliton}
When $k=1$ we have the three 3-partitions $(\tiny{\yng(1)}; \bullet; \bullet)$, $(\bullet; \tiny{\yng(1)}; \bullet)$, $(\bullet; \bullet; \tiny{\yng(1)})$ of $k$; our formula gives:
\begin{itemize}
\item Partition $(\tiny{\yng(1)}; \bullet; \bullet)$:
\begin{equation}
\begin{split}
& t^{-n+1}(1-t^{-1}) \left\langle W_{\square}^{U(3n)} \right \rangle \Big\vert_{(\tiny{\yng(1)}; \bullet; \bullet)} = \\
& \widetilde{\alpha}_1 (q-1)(1-t^{-1}) + \left(\widetilde{\alpha}_1 + \widetilde{\alpha}_2 + \widetilde{\alpha}_3 \right) \left(1-t^{-n}\right) \\
& - Q \widetilde{\alpha}_1 \dfrac{(1-q)(q-t)(1-t^{-1})(1-t^{-n})}{q^3(1-q^{-1}t^{1-n})}
\prod_{b=2}^3 \dfrac{(\widetilde{\alpha}_1 - t^{-n}\widetilde{\alpha}_b)(q\widetilde{\alpha}_1 - t\widetilde{\alpha}_b)}{(\widetilde{\alpha}_b - \widetilde{\alpha}_1)(\widetilde{\alpha}_1 - q^{-1}t^{1-n}\widetilde{\alpha}_b)} \\
& - Q \widetilde{\alpha}_1 t^{1-n} \dfrac{(1-q)(q-t)(1-t^{-1})(1-t^{1-n})(1-q^{-1}t^{-n})(1-q^{-2}t^{2-n})}{q^4(1-q^{-1}t^{1-n})(1-q^{-1}t^{2-n})(1-q^{-2}t^{1-n})} \\
& \times \prod_{b=2}^3 \dfrac{(\widetilde{\alpha}_b - t^{-n}\widetilde{\alpha}_1)(q\widetilde{\alpha}_b - t\widetilde{\alpha}_1)}{(\widetilde{\alpha}_1 - \widetilde{\alpha}_b)(\widetilde{\alpha}_b - q^{-1}t^{1-n}\widetilde{\alpha}_1)} \\
& - Q \sum_{a = 2}^3 \widetilde{\alpha}_a t^{1-n} \dfrac{(1-q)(q-t)(1-t^{-1})(1-t^{-n})}{q^4(1-q^{-1}t^{1-n})} \\
& \times \dfrac{(\widetilde{\alpha}_1 - t^{1-n}\widetilde{\alpha}_a)(\widetilde{\alpha}_1 - q^{-1} t^{-n}\widetilde{\alpha}_a)(\widetilde{\alpha}_1 - q^{-2}t^{2-n}\widetilde{\alpha}_a)(q \widetilde{\alpha}_1 - t \widetilde{\alpha}_a)}{(\widetilde{\alpha}_a - \widetilde{\alpha}_1)(\widetilde{\alpha}_1 - q^{-1}t^{1-n} \widetilde{\alpha}_a)(\widetilde{\alpha}_1 - q^{-1}t^{2-n} \widetilde{\alpha}_a)(\widetilde{\alpha}_1 - q^{-2}t^{1-n} \widetilde{\alpha}_a)} \\
& \times \prod_{\substack{b = 2 \\ b \neq a}}^3 
\dfrac{(q \widetilde{\alpha}_b - t \widetilde{\alpha}_a)(\widetilde{\alpha}_b - t^{-n} \widetilde{\alpha}_a)}{(\widetilde{\alpha}_a - \widetilde{\alpha}_b)(\widetilde{\alpha}_b - q^{-1}t^{1-n} \widetilde{\alpha}_a)} \\
& + o(Q^2) \,; \label{N3k100}
\end{split}
\end{equation}
in the limit $n \rightarrow \infty$, this reduces to
\begin{equation}
\widetilde{\alpha}_1 (q + t^{-1} - q t^{-1}) + \widetilde{\alpha}_2 + \widetilde{\alpha}_3 + Q \widetilde{\alpha}_1 \dfrac{(1-q)(1-t)(q-t)(q \widetilde{\alpha}_1 - t \widetilde{\alpha}_2)(q \widetilde{\alpha}_1 - t \widetilde{\alpha}_3)}{q^3 t (\widetilde{\alpha}_1 - \widetilde{\alpha}_2)(\widetilde{\alpha}_1 - \widetilde{\alpha}_3)} + o(Q^2)
\end{equation}
which matches the results in Sections \ref{oscN3k1} and \ref{ADHMN3k1} for $\widetilde{\alpha}_{1,2,3} = \alpha_{1,2,3}$ and $Q = p(qt^{-1})^3$.

\item Partition $(\bullet; \tiny{\yng(1)}; \bullet)$: 

Can be obtained from \eqref{N3k100} by permutation of the $\widetilde{\alpha}_a$ parameters. In the limit $n \rightarrow \infty$, this reduces to
\begin{equation}
\widetilde{\alpha}_2 (q + t^{-1} - q t^{-1}) + \widetilde{\alpha}_1 + \widetilde{\alpha}_3 + Q \widetilde{\alpha}_2 \dfrac{(1-q)(1-t)(q-t)(q \widetilde{\alpha}_2 - t \widetilde{\alpha}_1)(q \widetilde{\alpha}_2 - t \widetilde{\alpha}_3)}{q^3 t (\widetilde{\alpha}_2 - \widetilde{\alpha}_1)(\widetilde{\alpha}_2 - \widetilde{\alpha}_3)} + o(Q^2)
\end{equation}
which matches the results in Sections \ref{oscN3k1} and \ref{ADHMN3k1} for $\widetilde{\alpha}_{1,2,3} = \alpha_{1,2,3}$ and $Q = p(qt^{-1})^3$.

\item Partition $(\bullet; \bullet; \tiny{\yng(1)})$:

Can be obtained from \eqref{N3k100} by permutation of the $\widetilde{\alpha}_a$ parameters. In the limit $n \rightarrow \infty$, this reduces to
\begin{equation}
\widetilde{\alpha}_3 (q + t^{-1} - q t^{-1}) + \widetilde{\alpha}_1 + \widetilde{\alpha}_2 + Q \widetilde{\alpha}_3 \dfrac{(1-q)(1-t)(q-t)(q \widetilde{\alpha}_3 - t \widetilde{\alpha}_1)(q \widetilde{\alpha}_3 - t \widetilde{\alpha}_2)}{q^3 t (\widetilde{\alpha}_3 - \widetilde{\alpha}_1)(\widetilde{\alpha}_3 - \widetilde{\alpha}_2)} + o(Q^2)
\end{equation}
which matches the results in Sections \ref{oscN3k1} and \ref{ADHMN3k1} for $\widetilde{\alpha}_{1,2,3} = \alpha_{1,2,3}$ and $Q = p(qt^{-1})^3$.

\end{itemize}

\subsection{$\Delta\text{ILW}_4$ Spectrum from $\text{eRS}_4$ Spectrum} \label{Sec:sec5.3}

\subsubsection{Zero solitons}

For $k=0$ we only have the empty 4-partition $(\bullet; \bullet; \bullet\; \bullet)$ of $k$. From our formula we obtain
\begin{equation}
\begin{split}
& t^{-n+1}(1-t^{-1}) \left\langle W_{\square}^{U(4n)} \right \rangle \Big\vert_{(\bullet; \bullet; \bullet; \bullet)} = \left(\widetilde{\alpha}_1 + \widetilde{\alpha}_2 + \widetilde{\alpha}_3 + \widetilde{\alpha}_4 \right) \left(1-t^{-n}\right) \\
& + Q t^{-n} \dfrac{(1-q)(1-t)(q-t)\left(1-t^{-n}\right)}{q^5 \left(1- q^{-1} t^{1-n}\right)} 
\sum_{a=1}^4 \widetilde{\alpha}_a \prod_{\substack{b=1 \\ b \neq a}}^4 \dfrac{(t \widetilde{\alpha}_a - q \widetilde{\alpha}_b)(\widetilde{\alpha}_b - t^{-n} \widetilde{\alpha}_a)}{(\widetilde{\alpha}_a - \widetilde{\alpha}_b)(\widetilde{\alpha}_b - q^{-1} t^{1-n} \widetilde{\alpha}_a)} \\
& + o(Q^2)
\end{split}
\end{equation}
which, in the limit $n \rightarrow \infty$, reduces to
\begin{equation}
\widetilde{\alpha}_1 + \widetilde{\alpha}_2 + \widetilde{\alpha}_3 + \widetilde{\alpha}_4 + o(Q^2) \,.
\end{equation}
This immediately matches the results in Sections \ref{oscN4k0} and \ref{ADHMN4k0} for $\widetilde{\alpha}_{1,2,3,4} = \alpha_{1,2,3,4}$.

\subsubsection{One soliton}
When $k=1$ we have the four 4-partitions $(\tiny{\yng(1)}; \bullet; \bullet; \bullet)$, $(\bullet; \tiny{\yng(1)}; \bullet; \bullet)$, $(\bullet; \bullet; \tiny{\yng(1)}; \bullet)$, $(\bullet; \bullet; \bullet; \tiny{\yng(1)})$ of $k$; our formula gives:

\begin{itemize}
\item Partition $(\tiny{\yng(1)}; \bullet; \bullet; \bullet)$:

\begin{equation}
\begin{split}
& t^{-n+1}(1-t^{-1}) \left\langle W_{\square}^{U(4n)} \right \rangle \Big\vert_{(\tiny{\yng(1)}; \bullet; \bullet; \bullet)} = \\
& \widetilde{\alpha}_1 (q-1)(1-t^{-1}) + \left(\widetilde{\alpha}_1 + \widetilde{\alpha}_2 + \widetilde{\alpha}_3 + \widetilde{\alpha}_4 \right) \left(1-t^{-n}\right) \\
& + Q \widetilde{\alpha}_1 \dfrac{(1-q)(q-t)(1-t^{-1})(1-t^{-n})}{q^4(1-q^{-1}t^{1-n})}
\prod_{b=2}^4 \dfrac{(\widetilde{\alpha}_1 - t^{-n}\widetilde{\alpha}_b)(q\widetilde{\alpha}_1 - t\widetilde{\alpha}_b)}{(\widetilde{\alpha}_b - \widetilde{\alpha}_1)(\widetilde{\alpha}_1 - q^{-1}t^{1-n}\widetilde{\alpha}_b)} \\
& + Q \widetilde{\alpha}_1 t^{1-n} \dfrac{(1-q)(q-t)(1-t^{-1})(1-t^{1-n})(1-q^{-1}t^{-n})(1-q^{-2}t^{2-n})}{q^5(1-q^{-1}t^{1-n})(1-q^{-1}t^{2-n})(1-q^{-2}t^{1-n})} \\
& \times \prod_{b=2}^4 \dfrac{(\widetilde{\alpha}_b - t^{-n}\widetilde{\alpha}_1)(q\widetilde{\alpha}_b - t\widetilde{\alpha}_1)}{(\widetilde{\alpha}_1 - \widetilde{\alpha}_b)(\widetilde{\alpha}_b - q^{-1}t^{1-n}\widetilde{\alpha}_1)} \\
& + Q \sum_{a = 2}^4 \widetilde{\alpha}_a t^{1-n} \dfrac{(1-q)(q-t)(1-t^{-1})(1-t^{-n})}{q^5(1-q^{-1}t^{1-n})} \\
& \times \dfrac{(\widetilde{\alpha}_1 - t^{1-n}\widetilde{\alpha}_a)(\widetilde{\alpha}_1 - q^{-1} t^{-n}\widetilde{\alpha}_a)(\widetilde{\alpha}_1 - q^{-2}t^{2-n}\widetilde{\alpha}_a)(q \widetilde{\alpha}_1 - t \widetilde{\alpha}_a)}{(\widetilde{\alpha}_a - \widetilde{\alpha}_1)(\widetilde{\alpha}_1 - q^{-1}t^{1-n} \widetilde{\alpha}_a)(\widetilde{\alpha}_1 - q^{-1}t^{2-n} \widetilde{\alpha}_a)(\widetilde{\alpha}_1 - q^{-2}t^{1-n} \widetilde{\alpha}_a)} \\
& \times \prod_{\substack{b = 2 \\ b \neq a}}^4 
\dfrac{(q \widetilde{\alpha}_b - t \widetilde{\alpha}_a)(\widetilde{\alpha}_b - t^{-n} \widetilde{\alpha}_a)}{(\widetilde{\alpha}_a - \widetilde{\alpha}_b)(\widetilde{\alpha}_b - q^{-1}t^{1-n} \widetilde{\alpha}_a)} \\
& + o(Q^2) \,; \label{N4k1000}
\end{split}
\end{equation}
in the limit $n \rightarrow \infty$, this reduces to
\begin{equation} 
\begin{split}
& \widetilde{\alpha}_1 (q + t^{-1} - q t^{-1}) + \widetilde{\alpha}_2 + \widetilde{\alpha}_3 + \widetilde{\alpha}_4 \\
& + Q \widetilde{\alpha}_1 \dfrac{(1-q)(1-t)(q-t)(q \widetilde{\alpha}_1 - t \widetilde{\alpha}_2)(q \widetilde{\alpha}_1 - t \widetilde{\alpha}_3)(q \widetilde{\alpha}_1 - t \widetilde{\alpha}_4)}{q^4 t (\widetilde{\alpha}_1 - \widetilde{\alpha}_2)(\widetilde{\alpha}_1 - \widetilde{\alpha}_3)(\widetilde{\alpha}_1 - \widetilde{\alpha}_4)} + o(Q^2)
\end{split}
\end{equation}
which matches the results in Sections \ref{oscN4k1} and \ref{ADHMN4k1} for $\widetilde{\alpha}_{1,2,3,4} = \alpha_{1,2,3,4}$ and $Q = p(qt^{-1})^4$.

\item Partition $(\bullet; \tiny{\yng(1)}; \bullet; \bullet)$:

Can be obtained from \eqref{N4k1000} by permutation of the $\widetilde{\alpha}_a$ parameters. In the limit $n \rightarrow \infty$, this reduces to
\begin{equation} 
\begin{split}
& \widetilde{\alpha}_2 (q + t^{-1} - q t^{-1}) + \widetilde{\alpha}_1 + \widetilde{\alpha}_3 + \widetilde{\alpha}_4 \\
& + Q \widetilde{\alpha}_2 \dfrac{(1-q)(1-t)(q-t)(q \widetilde{\alpha}_2 - t \widetilde{\alpha}_1)(q \widetilde{\alpha}_2 - t \widetilde{\alpha}_3)(q \widetilde{\alpha}_2 - t \widetilde{\alpha}_4)}{q^4 t (\widetilde{\alpha}_2 - \widetilde{\alpha}_1)(\widetilde{\alpha}_2 - \widetilde{\alpha}_3)(\widetilde{\alpha}_2 - \widetilde{\alpha}_4)} + o(Q^2)
\end{split}
\end{equation}
which matches the results in Sections \ref{oscN4k1} and \ref{ADHMN4k1} for $\widetilde{\alpha}_{1,2,3,4} = \alpha_{1,2,3,4}$ and $Q = p(qt^{-1})^4$.

\item Partition $(\bullet; \bullet; \tiny{\yng(1)}; \bullet)$:

Can be obtained from \eqref{N4k1000} by permutation of the $\widetilde{\alpha}_a$ parameters. In the limit $n \rightarrow \infty$, this reduces to
\begin{equation} 
\begin{split}
& \widetilde{\alpha}_3 (q + t^{-1} - q t^{-1}) + \widetilde{\alpha}_1 + \widetilde{\alpha}_2 + \widetilde{\alpha}_4 \\
& + Q \widetilde{\alpha}_3 \dfrac{(1-q)(1-t)(q-t)(q \widetilde{\alpha}_3 - t \widetilde{\alpha}_1)(q \widetilde{\alpha}_3 - t \widetilde{\alpha}_2)(q \widetilde{\alpha}_3 - t \widetilde{\alpha}_4)}{q^4 t (\widetilde{\alpha}_3 - \widetilde{\alpha}_1)(\widetilde{\alpha}_3 - \widetilde{\alpha}_2)(\widetilde{\alpha}_3 - \widetilde{\alpha}_4)} + o(Q^2)
\end{split}
\end{equation}
which matches the results in Sections \ref{oscN4k1} and \ref{ADHMN4k1} for $\widetilde{\alpha}_{1,2,3,4} = \alpha_{1,2,3,4}$ and $Q = p(qt^{-1})^4$.

\item Partition $(\bullet; \bullet; \bullet; \tiny{\yng(1)})$:

Can be obtained from \eqref{N4k1000} by permutation of the $\widetilde{\alpha}_a$ parameters. In the limit $n \rightarrow \infty$, this reduces to
\begin{equation} 
\begin{split}
& \widetilde{\alpha}_4 (q + t^{-1} - q t^{-1}) + \widetilde{\alpha}_1 + \widetilde{\alpha}_2 + \widetilde{\alpha}_3 \\
& + Q \widetilde{\alpha}_4 \dfrac{(1-q)(1-t)(q-t)(q \widetilde{\alpha}_4 - t \widetilde{\alpha}_1)(q \widetilde{\alpha}_4 - t \widetilde{\alpha}_2)(q \widetilde{\alpha}_4 - t \widetilde{\alpha}_3)}{q^4 t (\widetilde{\alpha}_4 - \widetilde{\alpha}_1)(\widetilde{\alpha}_4 - \widetilde{\alpha}_2)(\widetilde{\alpha}_4 - \widetilde{\alpha}_3)} + o(Q^2)
\end{split}
\end{equation}
which matches the results in Sections \ref{oscN4k1} and \ref{ADHMN4k1} for $\widetilde{\alpha}_{1,2,3,4} = \alpha_{1,2,3,4}$ and $Q = p(qt^{-1})^4$.
\end{itemize}

\section{Comments on Quantum Cohomology of Instanton Moduli Spaces} \label{quantumcohomology}
As noticed in \cite{Feigin:2009ab} and further examined in \cite{Koroteev:2015dja}, the small $\gamma$ expansion of the first $\Delta\text{ILW}_1$ Hamiltonian $[\eta(z;p)]_1$ of Section \ref{Sec:sec2.2} contains the operator of quantum multiplication in the small quantum cohomology ring of the instanton moduli space $\mathcal{M}_{k,1}$ (i.e. Hilbert scheme of $k$ points on $\mathbb{C}^2$) introduced in  \cite{2004math.....11210O} and further discussed in \cite{2014JHEP...07..141B}. The claim is that the quantum multiplication operator \textit{almost} coincides with the first quantum $\text{ILW}_1$ Hamiltonian $\widehat{I}_3$, however, there are some subtleties which we shall address in this final section. Let us start with
\begin{equation}
\eta(z;p) = \text{exp}\left(\sum_{n>0}\lambda_{-n}z^n\right) \text{exp}\left(\sum_{n>0}\lambda_{n}z^{-n}\right)\,, 
\end{equation}
with commutation relations for the $\lambda_m$
\begin{equation}
[\lambda_m, \lambda_n] = -\dfrac{1}{m} \dfrac{(1-q^m)(1-t^{-m})(1-p^m)}{1-(pqt^{-1})^m} \delta_{m+n,0}\,.
\end{equation}
We now rewrite the $\lambda_m$ oscillators as
\begin{equation}
\lambda_m = \dfrac{1}{\vert m \vert} \sqrt{-\dfrac{(1-q^{\vert m \vert})(1-t^{-\vert m \vert})(1-p^{\vert m \vert})}{1-(pqt^{-1})^{\vert m \vert}}} \overline{a}_m\,, \label{lam}
\end{equation}
where the $\overline{a}_m$ oscillators satisfy commutation relations
\begin{equation}
[\overline{a}_m, \overline{a}_n] = m \delta_{m+n,0}\,.
\end{equation}
After substituting $p = -\tilde{p} /\sqrt{qt^{-1}}$\footnote{Remember that the quantum cohomology parameter is $\widetilde{p}$ and not $p$.} and defining $q=e^{i \gamma \epsilon_1}$, $t = e^{-i \gamma \epsilon_2}$ we can expand \eqref{lam} in small $\gamma$ and obtain
\begin{equation}
\begin{split}
\lambda_m \;=\; & \dfrac{1}{\vert m \vert} \sqrt{-\dfrac{(1-q^{\vert m \vert})(1-t^{-\vert m \vert})(1-(-\tilde{p}q^{-1/2}t^{1/2})^{\vert m \vert})}{1-(-\tilde{p}q^{1/2}t^{-1/2})^{\vert m \vert}}} \overline{a}_m \;= \\
=\; & \gamma \sqrt{\e_1 \e_2} \Bigg[ 1 + i \gamma \dfrac{\e_1 + \e_2}{4} m \dfrac{1 + (-\tilde{p})^m}{1 - (-\tilde{p})^m} + o(\gamma^2) \Bigg] \overline{a}_m \,.
\end{split}
\end{equation}
This leads to
\begin{equation}
[\eta(z;-\tilde{p}q^{-1/2}t^{1/2})]_1 = 1 + \gamma^2 \widehat{I}_2 + \gamma^3 \widehat{I}_3 + o(\gamma^4) \label{gf}
\end{equation}
where $\widehat{I}_2$ is the number operator
\begin{equation} 
\widehat{I}_2 = \e_1 \e_2 \sum_{m>0} \overline{a}_{-m}\overline{a}_m\,, 
\end{equation}
while
\begin{equation}
\widehat{I}_3 = i \e_1 \e_2  \dfrac{\epsilon_1 + \epsilon_2}{2} \sum_{m>0} m \dfrac{1 + (-\tilde{p})^m}{1 - (-\tilde{p})^m} \, \overline{a}_{-m} \overline{a}_m + \dfrac{(\e_1 \e_2 )^{\frac{3}{2}}}{2} \sum_{m,n > 0} (\overline{a}_{-m-n} \overline{a}_m \overline{a}_n + \overline{a}_{-m} \overline{a}_{-n} \overline{a}_{m+n})
\end{equation}
coincides with the first quantum $\text{ILW}_1$ Hamiltonian.
This is not completely unexpected: in fact, at least at the classical level, it is known that $\Delta\text{ILW}_1$ reduces to $\text{ILW}_1$ by taking an opportune $\gamma \rightarrow 0$ limit \cite{2009JPhA...42N4018S}. This limit is non-trivial since it requires to perform a Galilean transformation on the $\eta(z;p)$ field, so we cannot expect \eqref{gf} to be a generating function for the $\text{ILW}_1$ Hamiltonians\footnote{We thank Paolo Rossi for pointing this out.}. Let us also remark that in the limit $qt^{-1} = 1$ the Galilean transformation is trivial, the $\text{ILW}_1$ equation reduces to the Hopf (dispersionless KdV) equation, and \eqref{gf} reduces to the generating function for the quantum Hopf Hamiltonians studied in \cite{0036-0279-58-5-R03}. The same generating function appears in \cite{Eli,2008JGP....58..931R,2014arXiv1407.5824D} in relation to Symplectic Field Theory. However, the Galilean transformation only truly affects terms of order $o(\gamma^5)$, and this is why we obtain the first $\text{ILW}_1$ Hamiltonian with this method. 
We can now study eigenstates and eigenvalues for the Hamiltonian $\widehat{I}_3$ as we did in Section \ref{Sec:sec3}, by considering the eigenstate equation for states of level $k$ (eigenvalue of the number operator $\widehat{I}_2$). The eigenvalue $E^{(\lambda)}_3$ obtained in this way can unsurprisingly be recovered by expanding the $\Delta\text{ILW}_1$ eigenvalue $\mathcal{E}^{(1; \lambda)}_1$ in $\gamma$:
\begin{equation}
\mathcal{E}^{(1; \lambda)}_1 = 1 + \gamma^2 \e_1 \e_2 k + \gamma^3 E^{(\lambda)}_3 + o(\gamma^4) \,.
\end{equation}
We refer to \cite{Koroteev:2015dja} for further details. Here we want to remark that $\widehat{I}_3$ differs from the $\mathcal{M}_{k,1}$ operator of quantum multiplication of \cite{2004math.....11210O} by a term proportional to $\widehat{I}_2$. This additional term is related to the fact that for $\mathcal{M}_{k,1}$ an equivariant mirror map has to be taken into account in order to obtain the correct Gromov-Witten invariants, and is not present in the operators from \cite{2004math.....11210O}  for $\mathcal{M}_{k,N}$ when $N>1$, since the equivariant mirror map is trivial in these cases \cite{2014JHEP...01..038B}. In terms of integrable systems, this might be related to the Galilean transformation discussed above. We must admit that further study is required in order to clarify this possible equivalence. 

It is now not hard to imagine that the first $\Delta\text{ILW}_N$ Hamiltonian $\widehat{\mathcal{H}}^{(N)}_1$ and the first $\text{ILW}_N$ Hamiltonian $\widehat{I}_3^{(N)}$ will be related by an equation similar to \eqref{gf}, i.e.
\begin{equation}
\widehat{\mathcal{H}}^{(N)}_1 =  N + i \gamma \sum_{l=1}^N a_l + \gamma^2 \left( -\dfrac{1}{2}\sum_{l=1}^N a_l^2 + \widehat{I}_2^{(N)} \right) + \gamma^3 \left( -\dfrac{i}{6}\sum_{l=1}^N a_l^3 + \widehat{I}_3^{(N)} \right) + o(\gamma^4) \label{gfN} 
\end{equation}
with $\widehat{I}_2^{(N)}$ number operator
\begin{equation} 
\widehat{I}_2^{(N)} = \e_1 \e_2 \sum_{l=1}^N\sum_{m>0} \overline{a}_{-m}^{(l)}\overline{a}_m^{(l)}\,; 
\end{equation}
similarly, at the level of spectrum we expect
\begin{equation}
\mathcal{E}^{(N; \vec{\lambda})}_1 = N + i \gamma \sum_{l=1}^N a_l + \gamma^2 \left( -\dfrac{1}{2}\sum_{l=1}^N a_l^2 + \e_1 \e_2 k \right) + \gamma^3 \left( -\dfrac{i}{6}\sum_{l=1}^N a_l^3 + E^{(N; \vec{\lambda})}_3 \right) + o(\gamma^4) \,.
\end{equation}
For example, for $N = 2$ we obtain (for $\alpha_l = e^{i \gamma a_l}$)
\begin{equation}
\begin{split}
\widehat{I}_3^{(2)} & = i \e_1 \e_2  \dfrac{\epsilon_1 + \epsilon_2}{2} \sum_{m>0} m \dfrac{1 + \tilde{p}^m}{1 - \tilde{p}^m} \, \left( \overline{a}_{-m}^{(1)} \overline{a}_m^{(1)} + \overline{a}_{-m}^{(2)} \overline{a}_m^{(2)} \right) \\
& + i \e_1 \e_2 a_1 \sum_{m>0} \overline{a}_{-m}^{(1)}\overline{a}_m^{(1)} + i \e_1 \e_2 a_2 \sum_{m>0} \overline{a}_{-m}^{(2)}\overline{a}_m^{(2)} \\
& + i \e_1 \e_2 (\epsilon_1 + \epsilon_2) \sum_{m>0} \dfrac{m}{1-\widetilde{p}^m} \overline{a}_{-m}^{(1)}\overline{a}_m^{(2)} + i \e_1 \e_2 (\epsilon_1 + \epsilon_2) \sum_{m>0} \dfrac{m \widetilde{p}^m}{1-\widetilde{p}^m} \overline{a}_{-m}^{(2)}\overline{a}_m^{(1)} \\ 
& + \dfrac{(\e_1 \e_2 )^{\frac{3}{2}}}{2} \sum_{m,n > 0} \left(\overline{a}_{-m-n}^{(1)} \overline{a}_m^{(1)} \overline{a}_n^{(1)} + \overline{a}_{-m}^{(1)} \overline{a}_{-n}^{(1)} \overline{a}_{m+n}^{(1)} +
\overline{a}_{-m-n}^{(2)} \overline{a}_m^{(2)} \overline{a}_n^{(2)} + \overline{a}_{-m}^{(2)} \overline{a}_{-n}^{(2)} \overline{a}_{m+n}^{(2)} \right) \,,
\end{split}
\end{equation}
which can also be written as
\begin{equation}
\begin{split}
\widehat{I}_3^{(2)} & = i \e_1 \e_2  \dfrac{\epsilon_1 + \epsilon_2}{2} \sum_{m>0} m \left( \overline{a}_{-m}^{(1)} \overline{a}_m^{(1)} + \overline{a}_{-m}^{(2)} \overline{a}_m^{(2)} + 2 \, \overline{a}_{-m}^{(1)} \overline{a}_m^{(2)} \right) \\
& + i \e_1 \e_2 (\epsilon_1 + \epsilon_2) \sum_{m>0} \dfrac{m \widetilde{p}^m}{1-\widetilde{p}^m} 
\left( \overline{a}_{-m}^{(1)} \overline{a}_m^{(1)} + \overline{a}_{-m}^{(2)} \overline{a}_m^{(2)} + \overline{a}_{-m}^{(2)}\overline{a}_m^{(1)} + \overline{a}_{-m}^{(1)}\overline{a}_m^{(2)} \right) \\
& + i \e_1 \e_2 a_1 \sum_{m>0} \overline{a}_{-m}^{(1)}\overline{a}_m^{(1)} + i \e_1 \e_2 a_2 \sum_{m>0} \overline{a}_{-m}^{(2)}\overline{a}_m^{(2)} \\
& + \dfrac{(\e_1 \e_2 )^{\frac{3}{2}}}{2} \sum_{m,n > 0} \left(\overline{a}_{-m-n}^{(1)} \overline{a}_m^{(1)} \overline{a}_n^{(1)} + \overline{a}_{-m}^{(1)} \overline{a}_{-n}^{(1)} \overline{a}_{m+n}^{(1)} +
\overline{a}_{-m-n}^{(2)} \overline{a}_m^{(2)} \overline{a}_n^{(2)} + \overline{a}_{-m}^{(2)} \overline{a}_{-n}^{(2)} \overline{a}_{m+n}^{(2)} \right) \,.
\end{split}
\end{equation}
The above expression is precisely the operator of quantum multiplication in the small quantum cohomology ring of $\mathcal{M}_{k,2}$ given in \cite{2004math.....11210O} and it can be considered as a representation of Heisenberg algebra tensored with Virasoro algebra $\text{Heis}\otimes \text{Vir}$ \cite{2014JHEP...07..141B}. The same procedure works for all $N$ and the operator of quantum multiplication can be rewritten in terms of $\text{Heis}\otimes W_N$ algebra. 

The first quantum $\text{ILW}_2$ Hamiltonian, as well as few higher order Hamiltonians written as reperesentations of $\text{Heis}\otimes \text{Vir}$ have also been studied in detail in \cite{2013JHEP...11..155L}. It was noticed that when $\widetilde{p} \rightarrow 1$ the Heisenberg and the Virasoro parts decouple yielding the Hamiltonians of a free field and a dispersionless KdV system \`{a} la Bazhanov-Lukyanov-Zamolodchikov \cite{1996CMaPh.177..381B,1997CMaPh.190..247B,1999CMaPh.200..297B} respectively. One can imagine that for generic $N$ the Heisenberg part always decouples when $\widetilde{p} = 1$, leaving us with a $W_N$ algebra. 

This observation suggests that a similar factorization occurs at the level of q-$W_N$ algebras for $\widetilde{p} = 1$. One set of oscillators is expected to decouple, and the other oscillators can be used to construct a representation of the q-$W_N$ algebra \cite{1995q.alg.....7034S,1995q.alg.....8011A}.\footnote{Comments on the $\widetilde{p} = 1$ limit and the correspondence to q-$W_N$ can also be found in \cite{Feigin:2009ab}.}

\section{Future Directions}
In this work we have extended the duality between supersymmetric gauge theories and integrable systems to a new class of models by studying the large-$n$ limit of the five dimensional $\CN=1$ gauge theory ($\widehat{A}_0$-type) with matter in the adjoint representation. The next logical step should be to generalize our construction to affine quiver theories of higher rank (see \cite{Nekrasov:2012xe,Nekrasov:2013aa}) and investigate the variety of limits which appear inside each gauge group. We expect to thereby obtain a more generic effective theory which should describe a generalization of the ILW$_N$ family. For instance, the Seiberg-Witten curve for $\hat{A}_{N-1}$ theory in four dimensions describes the so-called elliptic spin-Calogero system, whose Hamiltonian also contains spin operators. When $N=1$ the spin part of the Hamiltonian reduces to the value of its Casimir. If each group inside the necklace $\hat{A}_{N-1}$ quiver is $U(n)$ one may ask what happens along the lines of this paper with the theory and with its dual integrable system. 

We have mentioned that we do not possess a compact form for RS$_N$ Hamiltonians. We hope that studying gauge theories with defect as in \cite{Bullimore:2015fr} for more generic theories will help to obtain those operators and prove, using the recent approach of \cite{Nekrasov:2015wsu} that the supersymmetric partition functions and BPS observables of those theories provide a formal solution to the dual integrable models.

We hope to find a relationship between the two different elliptic deformations of Ding-Iohara algebra provided in \cite{Feigin:2009ab} and in \cite{2013arXiv1301.4912S}. A better understanding of this point will probably clarify the connection between the two algebras, the gauge theories studied in this paper, and the elliptic Virasoro algebra introduced in \cite{2015arXiv151100574N,2015arXiv151100458I}. Additionally we expect to get a deeper understanding of the AGT correspondence including its difference and elliptic versions.

In addition five dimensional theories may have Chern-Simons terms. One should be able to study their imprint on the large-$n$ physics which have been discussed in this paper. Recently there has been progress in understanding 5d/3d systems in the presence of Chern-Simons terms \cite{Aganagic:2015aa,Aganagic:2014ab,Aganagic:2014aa}. See also \cite{Gorsky:2015aa} for the discussion of large-$n$ transitions for theories with fractional Chern-Simons terms.

\section*{Acknowledgements}
We would like to thank Sergei Gukov, Paolo Rossi and Fabrizio Nieri for helpful discussions.
PK would also like to thank Korean Institute for Advanced Study, California Institute of Technology, Kavli Institute for Physics and Mathematics of the Universe, and W. Fine Institute for Theoretical Physics at University of Minnesota for kind hospitality during his visits, where part of his work was done.
The research of PK was supported by the Perimeter Institute for Theoretical Physics. Research at Perimeter Institute is supported by the Government of Canada through Industry Canada and by the Province of Ontario through the Ministry of Economic Development and Innovation.

\newpage

\appendix

\section{The ADHM quiver and Bethe Ansatz Equations for $\text{ILW}_N$} \label{appA}

In this Appendix we will consider the $\mathcal{N} = 2^*$ ADHM quiver theory on $\mathbb{CP}^1 \times S^1_{\gamma}$ inside the 11d geometry $\Complex_q \times \Complex_t \times \Complex \times \mathcal{O}(-2)_{\mathbb{CP}^1} \times S^1_{\gamma}$. The field content of the quiver is given as follows:

\begin{table}[h!]
\begin{center}
\begin{tabular}{c|c|c|c|c|c}
{} & $\chi$ & $B_{1}$ & $B_{2}$ & $I$ & $J$ \\ \hline
D-brane sector & D2/D2 & D2/D2 & D2/D2 & D2/D6 & D6/D2 \\ \hline
gauge $U(k)$ & $Adj$ & $Adj$ & $Adj$ & $\mathbf{k}$ & $\mathbf{\bar{k}}$ \\ \hline
flavor $U(N)\times U(1)^{2}$ & $\mathbf{1}_{(-1,-1)}$ & $\mathbf{1}_{(1,0)}$ & $\mathbf{1}_{(0,1)}$ & $\mathbf{\bar{N}}_{(0,0)}$ & $\mathbf{N}_{(1,1)}$ \\ \hline
twisted masses & $\epsilon_1 + \epsilon_2$ & $-\epsilon_{1}$ & $-\epsilon_{2}$ & $-a_{j}$ & $a_{j}-\epsilon_1 - \epsilon_2$ \\ \hline
$R$-charge & $2$ & $0$ & $0$ & $0$ & $0$ \\ \hline
\end{tabular} 
\caption{Matter content of the ADHM Gauged Linear Sigma Model.}
\end{center}
\end{table} 

\begin{figure}[h!]
  \centering
\includegraphics[width=0.6\linewidth]{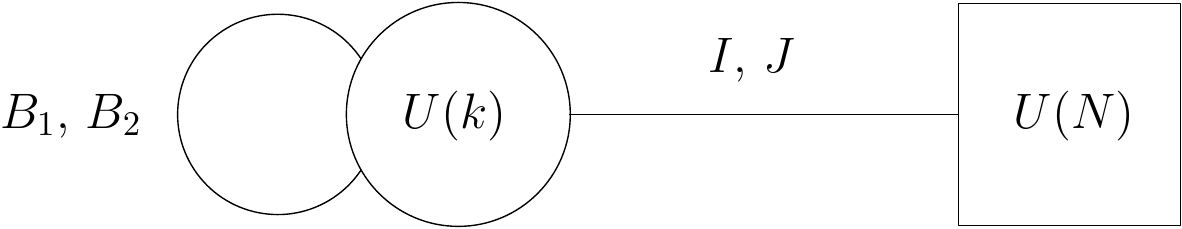} 
  \caption{The ADHM quiver.\label{fig:A}}
\end{figure}

\noindent The superpotential is given by $W=\textrm{Tr}_{k}\left\{\chi\left([B_{1},B_{2}]+IJ\right)\right\}$.
When $\epsilon_1 + \epsilon_2 = 0$ the $\mathcal{N}=2$ vector supermultiplet and the $\mathcal{N}=2$ adjoint chiral supermultiplet $\chi$ combine into an $\mathcal{N}=4$ vector supermultiplet, while in the $\epsilon_1 + \epsilon_2 \neq 0$ case supersymmetry is broken to $\mathcal{N} = 2^*$.
The moduli space $\mathcal{M}_{k,N}$ of supersymmetric vacua in the Higgs branch is obtained by setting to zero the VEV of the adjoint scalar field in the $\chi$ supermultiplet and it is given by the solutions of the $F$ and $D-$term equations modulo the action of the gauge group $U(k)$: \vspace*{0.3 cm}
\begin{center}
\begin{tabular}{l}
$\mathcal{M}_{k,N}$ = \Bigg\{ 
\begin{tabular}{ll}
$[B_{1},B_{2}] + IJ = 0$ & ($F$-term) \\ 
$[B_{1},B_{1}^{\dagger}] + [B_{2},B_{2}^{\dagger}] + I I^{\dagger} -J^{\dagger}J = \xi$ & ($D$-term) \\ 
\end{tabular} 
\Bigg\} \Bigg/ $U(k)$\,,
\end{tabular} 
\end{center} \vspace*{0.3 cm}
where $\xi$ is Fayet-Iliopoulos parameter. This manifold can be identified with the ADHM moduli space of $k$ instantons for a pure $U(N)$ Yang-Mills theory. In terms of a D2/D6 brane system, the $k$ D2 branes wrapped on $\mathbb{P}^1 \times S^1_{\gamma}$ can be understood as a $k$-instanton configuration for the pure $U(N)$ supersymmetric theory living on the $N$ D6 branes wrapping $\Complex_q \times \Complex_t \times \mathbb{P}^1 \times S^1_{\gamma}$ (here $q = e^{i \gamma \epsilon_1}$, $t = e^{-i \gamma \epsilon_2}$). As is well known in the context of D(p-4)/Dp brane systems, the auxiliary 3d theory living on D2 branes is precisely the ADHM quiver theory whose Higgs branch vacua describes $\mathcal{M}_{k,N}$. When the radius of the $S^1_{\gamma}$ circle is sent to zero we go back to the setting of \cite{2014JHEP...07..141B} with a system of $k$ D1 branes and $N$ D5 branes wrapping respectively $\mathbb{P}^1$ and $\Complex_q \times \Complex_t \times \mathbb{P}^1 $ inside the 10d geometry $\Complex_q \times \Complex_t \times \Complex \times \mathcal{O}(-2)_{\mathbb{P}^1}$.
 
One knows from the Bethe/Gauge correspondence \cite{Nekrasov:2009uh,Nekrasov:2009ui} that the equations determining the Coulomb branch vacua coincide with Bethe Ansatz Equations for a quantum spin chain 
\begin{equation}
\begin{split}
& 
\prod_{l=1}^N \sin [\frac{\gamma}{2} (\Sigma_s - a_l)]
\prod_{\substack{t = 1 \\ t\neq s}}^k \dfrac{\sin [\frac{\gamma}{2} (\Sigma_{st} - \epsilon_1)] \sin [\frac{\gamma}{2} (\Sigma_{st} - \epsilon_2)]}{\sin [\frac{\gamma}{2} ( \Sigma_{st})] \sin [\frac{\gamma}{2} (\Sigma_{st} - \epsilon)]} = \\
& e^{-2 \pi \xi}\, \prod_{l=1}^N \sin [\frac{\gamma}{2} ( -\Sigma_s + a_l - \epsilon)] 
\prod_{\substack{t = 1 \\ t\neq s}}^k \dfrac{\sin [\frac{\gamma}{2} ( \Sigma_{st} + \epsilon_1)] \sin [\frac{\gamma}{2} (\Sigma_{st} + \epsilon_2)]}{\sin [\frac{\gamma}{2} ( \Sigma_{st})] \sin [\frac{\gamma}{2} (\Sigma_{st} + \epsilon)]} \,. \label{BAEapp}
\end{split}
\end{equation}
Here $\epsilon = \epsilon_1 + \epsilon_2$ and $\Sigma_s$ are the scalars in the 2d $\mathcal{N} = (2,2)$ superfield strength multiplet arising when $S^1_{\gamma}$ shrinks to zero size; the effect of the finite-size $S^1_{\gamma}$ circle consists in having to take into account all the Kaluza-Klein modes, which generate the sine functions in \eqref{BAEapp}. 

When $\xi \rightarrow \infty$, the solutions to \eqref{BAEapp} are labelled by $N$-partitions $\vec{\lambda} = (\lambda^{(1)}; \ldots; \lambda^{(N)})$ of $k$; in the simplest example of $N=1$ these are given by (setting $a_1 = 0$)
\begin{equation}
\Sigma_s = (i-1)\e_1 + (j-1)\e_2 \;\;\;\text{mod } 2 \pi i 
\end{equation} 
with $i$, $j$ running over the boxes of the partition $\lambda^{(1)} = \lambda$, where in the general case we will have
\begin{equation}
\Sigma_s^{(l)} = a_l + (i-1)\e_1 + (j-1)\e_2 \;\;\;\text{mod } 2 \pi i \;\;\;,\;\;\; l = 1, \ldots, N
\end{equation} 
with $i$, $j$ running over the boxes of the partition $\lambda^{(l)}$. \\
For $\xi$ finite we can define $\sigma_s = e^{i \gamma \Sigma_s}$, $q = e^{i \gamma \epsilon_1}$, $t = e^{-i \gamma \epsilon_2}$, $\alpha_l = e^{i \gamma a_l}$ and rewrite \eqref{BAEapp} as 
\begin{equation}
\begin{split}
& \prod_{l=1}^N (\sigma_s \alpha_l^{-1} - 1) \prod_{\substack{t = 1 \\ t\neq s}}^k \dfrac{(\sigma_{s} - q \sigma_t) (\sigma_{s} - t^{-1} \sigma_t)}{( \sigma_{s} - \sigma_t) (\sigma_{s} - q t^{-1} \sigma_{t})} = \\
& = \widetilde{p} \, (-\sqrt{q t^{-1}})^N \, \prod_{l=1}^N (\sigma_s \alpha_l^{-1} - q^{-1} t) \prod_{\substack{t = 1 \\ t\neq s}}^k \dfrac{(\sigma_{s} - q^{-1} \sigma_t) (\sigma_{s} - t \sigma_t)}{(\sigma_{s} - \sigma_t) (\sigma_{s} - q^{-1} t \sigma_{t})}\,. \label{BAE2app}
\end{split}
\end{equation}
Our results in the main text give strong evidence that these equations coincide describe spectrum of quantum $\Delta\text{ILW}_N$. Perturbatively in $e^{-2\pi\xi}$ the solutions to \eqref{BAE2app} are still labelled by $N$-partitions $\vec{\lambda}$ of $k$, and the eigenvalue of the first $\Delta\text{ILW}_N$ Hamiltonian is given by
\begin{equation}
\begin{split}
\mathcal{E}^{(N; \vec{\lambda})}_1 & = \sum_{l=1}^N \left[ \alpha_l - (1-q)(1-t^{-1}) \sum_s \sigma_s^{(l)} \Big\vert_{\lambda^{(l)}} \right] \\
& = \sum_{l=1}^N \alpha_l - (1-q)(1-t^{-1}) \sum_s \sigma_s \Big\vert_{\vec{\lambda}}  \label{chernapp}
\end{split}
\end{equation}
where $\text{Tr}\sigma = \sum_s \sigma_s = \sum_s e^{i \gamma \Sigma_s}$ is evaluated at the solutions of \eqref{BAE2app}. In the $\Delta\text{BO}_N$ limit \eqref{chernapp} reduces to
\begin{equation}
\begin{split}
\mathcal{E}^{(N; \vec{\lambda})}_1 &= \sum_{l=1}^N \alpha_l + (1-t^{-1}) \sum_{l=1}^N \sum_{j=1}^k \alpha_l (q^{\lambda_j^{(l)}} - 1)t^{1-j} \\ 
& = \sum_{l=1}^N \alpha_l - (1-q)(1-t^{-1})\sum_{l=1}^N \alpha_l \sum_{(i,j) \in \lambda^{(l)}} q^{i-1} t^{1-j} \,. \label{appDBON} 
\end{split}
\end{equation}

\bibliography{cpn1}
\bibliographystyle{JHEP}

\end{document}